\documentclass[11pt,a4paper]{article}
\usepackage[square, super]{natbib}

\usepackage[utf8]{inputenc}
\usepackage{amsmath}
\usepackage{authblk}
\usepackage{graphicx}
\usepackage[margin=1in]{geometry}
\usepackage{textcomp}
\usepackage[T1]{fontenc}
\usepackage{authblk}
\usepackage{hyperref}
\usepackage{array}
\usepackage{multirow} % 引入multirow包
\usepackage{tabularx} 
\usepackage{indentfirst}
\usepackage{subfigure}
\usepackage{diagbox}
\usepackage{booktabs}
\usepackage{float}
\usepackage{bm}
\usepackage{adjustbox}
\usepackage{threeparttable}

\newcommand\keywords[1]{\textbf{Keywords}: #1}
\title{A new solar radiation pressure model for some orbit types in the cislunar space}
\author[1,2,3]{Haohan Li \thanks{First author: email: lihh@smail.nju.edu.cn}}
\author[1,2,3]{Yuxuan Miao}
\author[1,2,3*]{Xiyun Hou \thanks{Corresponding author: email: houxiyun@nju.edu.cn}}
\author[4]{Jinjun Zheng}
\author[4]{Xiangjun Wu}
\author[4]{Haihong Wang}
\author[4]{Guodong Zhang}

\affil[1] {School of Astronomy and Space Science, Nanjing University, Nanjing 210023, China}
\affil[2]  {Institute of Space Environment and Astrodynamics, Nanjing University, Nanjing 210023, China}
\affil[3] {Key Laboratory of Modern Astronomy and Astrophysics, Ministry of Education, Nanjing 210023, China}
\affil[4] {Institute of Telecommunication and Navigation Satellites,China Academy of Space Technology,Beijing 100094,China}
\date{}

\begin{document}
\maketitle

\begin{abstract}
%%%
For satellites in the cislunar space, solar radiation pressure (SRP) is the third largest perturbation, which is only less significant than the lunisolar gravity perturbations. It is the primary factor limiting the accuracy of orbit determination for such satellites. Up to now, numerous SRP models have been proposed for artificial satellites close to the Earth, but these models have their shortcomings when applied to satellites in the cislunar space. In this study, we concentrate on various scenarios of cislunar satellites in periodic or quasi-periodic orbits. We first employ the box-wing model to simulate the SRP effects and then propose an appropriate general SRP model based on these simulations, termed Empirical NJU Cislunar Model (ENCM). Additionally, several scenario-specific sub-models suited to different mission profiles are developed. Furthermore, the proposed model is verified in the orbit determination process. Comparisons with the conventional cannonball and ECOM models demonstrate that the ENCM model yields a significant improvement in orbit determination accuracy, showing promising potential for future cislunar missions.
%%%%
\end{abstract}


\section{Introduction}

\par Due to the proliferation of space debris and the growing number of small satellites, the near-Earth space (in this work, it means the region lower than the geosynchronous orbit (GEO) altitude) is becoming increasingly congested. Concurrently, key technologies for lunar-distance operations, including satellite launch, communication/navigation (C/N), and control have achieved significant progresses. The advantages of cislunar space, compared to the limitations of near-Earth space, are garnering considerable interest from the space community. Space agencies and companies are making their plans utilizing the whole cislunar space \cite{WH_2022}. This trend underscores the need to enhance humanity's current C/N and space situational awareness (SSA) capabilities to accommodate the needs of the cislunar space. Orbit determination (OD) is a cornerstone of these capabilities. For satellites operating approximately at the distance of the Moon but not very close to it, such as the distant retrograde orbits (DROs) \cite{Liu_2014}, the nearly rectilinear halo orbits (NRHOs) \cite{mccarthy_2021}, and the triangular libration points orbits (TLPOs) \cite{hou_2015}, the SRP is the third largest perturbation \cite{hou_2024} and cannot be neglected in the orbit determination (OD) process. Nevertheless, accurately modelling the SRP is not an easy task because this effect depends critically on the target's geometry, attitude, surface material, and the solar radiation flux. In general, SRP models can be categorized into three types: analytical models, semi-analytical models, and empirical models.
\par Extensive research has been conducted on the SRP model for Earth-orbiting satellites. Among these models, the cannonball model is the most basic and straightforward, often serving the purpose of a preliminary analysis \citep{yin_2024,li_2023} or the situation in which accurate modeling of the SRP is not required \cite{liu_2024}. Nonetheless, its applicability is limited to the assumption of a spherical satellite, such as LAGEOS. The box-wing model is another type of analytical SRP model, offering a more sophisticated approach to model the SRP. It divides the satellite into a main body (modelled as a box) and two solar panels (modelled as wings), allowing for individual calculation of the SRP on each surface of the box and the two wings. When the relative orientation to the Sun and the properties of each surface are known, this model can provide a precise depiction of the SRP effects. For satellites with sophisticated shapes, the box-wing model can evolve into the N-plane model \cite{hesar_2016,mao_2024}. Furthermore, dedicated analytical SRP models can be derived from specific satellites. The Rock model is constructed by Fliegel et al. \cite{fliegel_1992,fliegel_1996} to model the SRP of the GPS Block I, Block II, and Block IIR satellites. In this model, SRP is decomposed into three orthogonal components in the satellite body frame. This model can produce a reasonable prior estimation of the true SRP. For more complex shaped satellites, pixel array is an effective way to develop a more elaborate model. Ziebart and Dare consider the complex surface model and analyse the SRP effects for GLONASS by using the pixel array \cite{ziebart_2001}. Due to the complex space environment, the physical properties of the surface may change over time. Thus, parameters of analytical models may also change. It means that we need to recover these parameters in the OD process. Building upon the box-wing model, Rodriguez-Solano et al. proposed a modified box-wing model, called adjusted box-wing model, which is a kind of semi-analytical SRP model \cite{rodriguez_2012}. This model redefines nine optical and attitude parameters to simplify the OD process. Both analytical and semi-analytical models require the information of real-time satellite attitude. As a result, they are not practical for satellites without attitude information.
\par  In the OD process, empirical models play a pivotal role, relying fundamentally on basic analysis of SRP frequencies and are characterized by a relatively low number of parameters. Among these models, the Empirical CODE Orbit Model (ECOM) stands out as the most prevalent one, which was developed in the early 1990s by the Center for Orbit Determination in Europe (CODE), specifically for GPS satellites \cite{beutler_1994,springer_1999}. However, when the ECOM is applied to the GLONASS satellites, problems of the ECOM appear due to the the odd draconitic harmonics in GNSS geodetic products \cite{rodriguez_2014} and these problems are in line with the conclusion in \cite{MEINDL_2013}. In 2014, Arnold et al. introduced another empirically derived model, termed ECOM2 model \cite{arnold_2015}, which has gained widespread application and has been demonstrated to be compatible with both the Galileo and the QZSS satellites \cite{prange_2017}. Both ECOM and ECOM2 models have also been utilized by other GNSS constellations, such as the BeiDou Navigation Satellite System (BDS) \cite{guo_2017,liu_2019}. Nevertheless, for satellites in the cislunar space, no commonly adopted empirical SRP model yet exists.  
\par There are many different types of orbits in the cislunar space. Since these orbits are typically at considerable distances from Earth, the SRP's influence is significant, making accurate modeling of the SRP essential for the OD accuracy. This study focuses on constructing appropriate SRP models for satellites in periodic and quasi-periodic orbits within the Earth-Moon synodic frame—specifically, distant retrograde orbits (DROs), nearly rectilinear halo orbits (NRHOs), and quasi-periodic orbits around the triangular libration points. Design methods and characteristics of these orbits in the realistic Earth-Moon system are detailed in related studies \cite{hou_2010,liu_2024_stability,sun_2024,browna_2024}. A general new SRP model termed Empirical NJU Cislunar Model (ENCM) is proposed for the OD of these cislunar satellites, and several sub-models are developed for different scenarios and different orbit types. The use of the ENCM is shown to significantly improve the OD accuracy when compared to results of either the cannonball model or the ECOM.
\par The remainder of this paper is organized as follows. Basic theory regarding the SRP and the construction of the ENCM are introduced in Section 2. Section 3 constructs the general ENCM model and compares it with reference SRP forces computed using the box-wing model. Section 4 applies the new SRP models to the OD process and evaluates the improvement in accuracy by comparing the results with those from other models. Section 5 discusses the simplified sub-models of ENCM in different scenarios and several challenges, while section 6 concludes the whole work.

\section{Theoretical SRP model}

\subsection{Basic SRP model}
\par For a surface composed of the same material, the SRP can be described using three optical coefficients: the absorptivity $\alpha$, the specular reflectivity $\rho$, and the diffuse reflectivity $\delta$. All coefficients range from 0 to 1, and satisfy the following relationship \cite{hesar_2016}:
\begin{equation}
    \alpha+\rho+\delta=1.
\end{equation}
In Fig. 1, two directions and an angle are defined: the direction opposite to the incident ray $\mathbf{e}_S$, the outward normal direction of the plane $\mathbf{e}_N$, and the angle $\theta$ between these two directions.
\begin{figure}[htp]
    \centering
    \includegraphics[width=8cm]{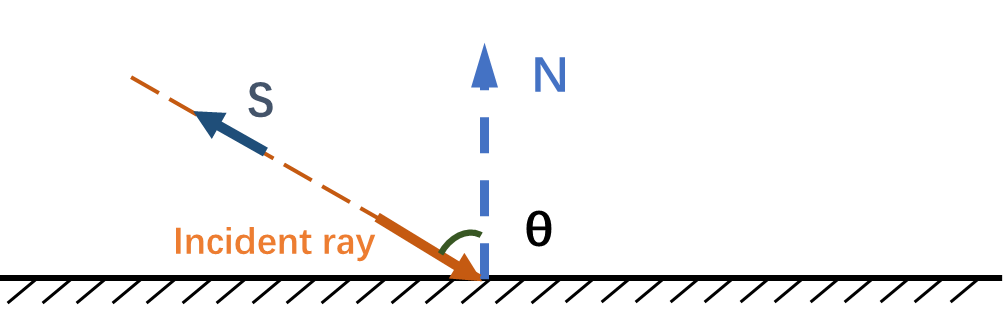}
    \caption{Light ray incidents on a plane}
\end{figure}
\par The radiation force corresponding to these three optical processes is given respectively by \cite{hesar_2016}:
\begin{itemize}
    \item [(1)] Force due to absorption $\mathbf{F}_{\alpha}$
    \begin{equation}
        \mathbf{F}_{\alpha}=-\frac{PS\cos{\theta}}{c}\alpha\cdot\mathbf{e}_S.\nonumber
    \end{equation}
    \item [(2)] Force due to specular reflection $\mathbf{F}_{s}$
    \begin{equation}
        \mathbf{F}_{s}=-2\frac{PS\cos{\theta}}{c}\rho\cos{\theta}\cdot\mathbf{e}_N.\nonumber
    \end{equation}
    \item [(3)] Force due to diffuse reflection $\mathbf{F}_{d}$
    \begin{equation}
        \mathbf{F}_{d}=-\frac{PS\cos{\theta}}{c}\delta(\mathbf{e}_S+B\cdot\mathbf{e}_N).\nonumber
    \end{equation}
\end{itemize}
 Here, $P=P_0\times{D^2}/{r^2}$ is the solar radiation pressure at distance $r$; $r$ is the distance from the Sun to the surface element; $D$ is average Sun-Earth distance (1AU); $P_0$ is the solar radiation flux at the distance $D$; and $B$ is a diffuse reflection coefficient. For an ideal Lambertian surface, $B = 2/3$ \cite{fliegel_1992}. Therefore, the total SRP of a planar surface is the sum of these components:
\begin{equation}
\begin{aligned}
    \mathbf{F}_{SRP}&=\mathbf{F}_{\alpha}+\mathbf{F}_{s}+
    \mathbf{F}_{d}
    \\&=-\frac{PS\cos{\theta}}{c}
    [(\alpha+\delta)\cdot\mathbf{e}_S+(2\rho\cos{\theta}+\frac{2}{3}\delta)\cdot\mathbf{e}_N].
\end{aligned}
\end{equation}
\\ For a cylindrical surface, this force model has been formulated by Fliegel \cite{fliegel_1992} as:
\begin{equation}
    \mathbf{F}_{SRP}=-\frac{PS\cos{\theta}}{c}
    [(\alpha+\delta)\cdot\mathbf{e}_S+(\frac{4}{3}\rho\cos{\theta}+\frac{\pi}{6}\delta)\cdot\mathbf{e}_N].
\end{equation}
\par In more realistic situations, the SRP force is often decomposed into three orthogonal components within a specified coordinate system. A commonly used system is the DYB coordinate frame. Following Arnold's definition for the ECOM2 model \cite{arnold_2015}, the DYB frame is constructed as follows \cite{arnold_2015}: Denote $\mathbf{r}_{p}$ as the vector from the Earth to the satellite, $\mathbf{r}_{sp}$ as the vector from the Sun to the satellite, and $\mathbf{e}_r$ as the unit vector of $\mathbf{r}_p$. The DYB frame can be defined as \cite{arnold_2015}:
\begin{equation}
    \mathbf{e}_{D}=-\frac{\mathbf{r}_{sp}}{|{r}_{sp}|} \qquad
    \mathbf{e}_{Y}=\frac{\mathbf{e}_{D}\times\mathbf{e}_{r}}{|\mathbf{e}_{D}\times\mathbf{e}_{r}|} \qquad
    \mathbf{e}_{B}=\frac{\mathbf{e}_{D}\times\mathbf{e}_{Y}}{|\mathbf{e}_{D}\times\mathbf{e}_{Y}|}.
\end{equation}
\begin{figure}[htp]
    \centering
    \includegraphics[width=6cm]{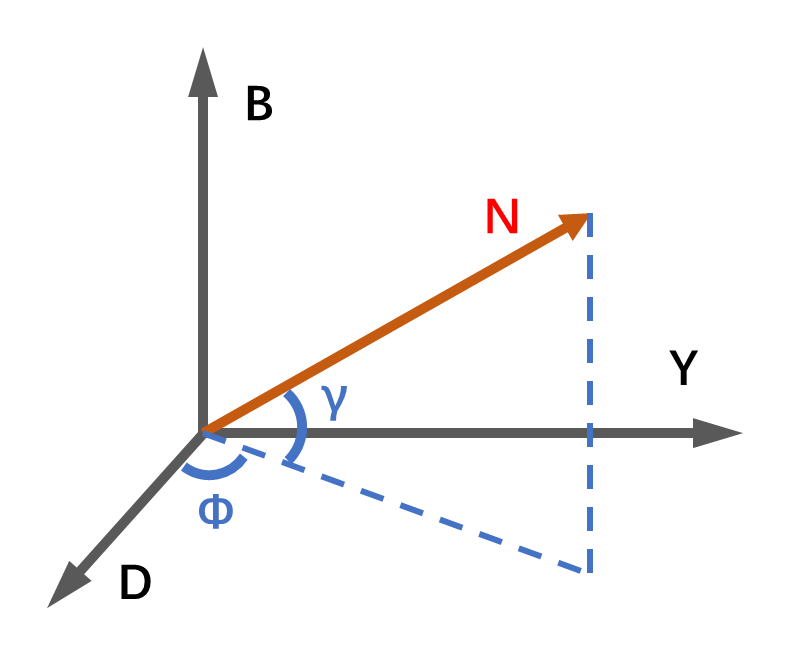}
    \caption{The relationship between $\mathbf{N}$ and DYB}
\end{figure}
For a planar surface element, we can use two angles to describe the normal direction $\mathbf{e}_N$, as shown in Fig. 2. After some derivations, we have 
\begin{equation}
\begin{aligned}
    &\mathbf{F}_{D}=-\frac{{F}_{0}}{{r}^{2}}(P_{1}\cos\phi\cos\gamma+P_{2}\cos^{2}{\phi}\cos^{2}{\gamma}+P_{3}\cos^{3}{\phi}\cos^{3}{\gamma})\mathbf{e}_D,\\
    &\mathbf{F}_{Y}=-\frac{{F}_{0}}{{r}^{2}}(P_{2}\cos{\phi}\sin{\phi}\cos^{2}\gamma+P_{3}\cos^{2}{\phi}\sin{\phi}\cos^{3}{\gamma})\mathbf{e}_Y,\\
    &\mathbf{F}_{B}=-\frac{{F}_{0}}{{r}^{2}}(P_{2}\cos{\phi}\cos{\gamma}\sin{\gamma}+P_{3}\cos^{2}{\phi}\cos^{2}{\gamma}\sin{\gamma})\mathbf{e}_B.
\end{aligned}
\end{equation}
\\ where ${F}_{0}=P_{0}S/(mc)$; $P_1=\alpha+\delta$; $P_2=B\delta$; and $P_3=2\rho$. From the equations above, it is obvious that the SRP only depends on three factors: 
\begin{itemize}
    \item [(1)] the surface property coefficients ($\alpha$, $\rho$, $\delta$)
    \item [(2)] the distance from the Sun to the satellite
    \item [(3)] the direction of the normal vector
\end{itemize}
\par It is usually reasonable to assume that the surface optical properties remain constant over short time spans. Furthermore, for a satellite with a fixed attitude relative to an inertial frame or a target, factors (2) and (3) are governed predominantly by the satellite's orbital frequencies. Consequently, the simplified SRP model can be formulated as a function of the orbital frequencies or the corresponding angles.

\subsection{The Empirical CODE Orbit Model (ECOM)}
\par ECOM is one of the most widely used empirical SRP models for the GNSS satellites. It can take different forms depending on specific satellites and application scenarios. All these forms are based on the DYB coordinate and decompose the SRP into three orthogonal directions. Therefore, the SRP can be expressed as \cite{fliegel_1992}:
\begin{equation}
\mathbf{a}_{SRP}=D\cdot\mathbf{e}_{D}+Y\cdot\mathbf{e}_{Y}+B\cdot\mathbf{e}_{B}.
\nonumber
\end{equation}
In most cases, a priori estimation of the SRP (denoted as $\mathbf{a}_{0}$) is available, so the expression is modified as
\begin{equation}
\mathbf{a}_{SRP}=\mathbf{a}_{0}+D\cdot\mathbf{e}_{D}+Y\cdot\mathbf{e}_{Y}+B\cdot\mathbf{e}_{B}.
\nonumber
\end{equation}
In the above equation, $D$, $Y$, and $B$ are the coefficients along the three orthogonal directions. Different types of SRP models define the DYB coordinate system in different ways and have different expressions of these coefficients. In ECOM, the DYB coordinate system is defined as \cite{fliegel_1996}:
\begin{equation}
    \mathbf{e}_{D}=\frac{\mathbf{r}_{sp}}{|{r}_{sp}|}; \qquad
    \mathbf{e}_{B}=\frac{\mathbf{e}_{D}\times\mathbf{e}_{Y}}{|\mathbf{e}_{D}\times\mathbf{e}_{Y}|}.
    \nonumber
\end{equation}
where $\mathbf{r}_{sp}$ is the vector from the Sun to the satellite, and $\mathbf{e}_{Y}$ is the unit vector along the satellite's panel axis \cite{fliegel_1992}. The model can be expressed as trigonometric functions of angle $u$ and the nine parameters \cite{fliegel_1992}:
\begin{equation}
\begin{aligned}\left\{\begin{aligned}
    {D}(u)&={D}_{0}+{D}_{1}\cos{u}+{D}_{2}\sin{u},\\
    {Y}(u)&={Y}_{0}+{Y}_{1}\cos{u}+{Y}_{2}\sin{u},\\
    {B}(u)&={B}_{0}+{B}_{1}\cos{u}+{B}_{2}\sin{u}.
\end{aligned}\right.\end{aligned}
\end{equation}
where $u=f+\omega$ is the orbital latitude of the satellite. A simplified version of the model reduces the nine parameters into five parameters, in the form of \cite{fliegel_1996}:
\begin{equation}
\begin{aligned}\left\{\begin{aligned}
    {D}(u)&={D}_{0}+{D}_{1}\cos{u}+{D}_{2}\sin{u},\\
    {Y}(u)&={Y}_{0},\\
    {B}(u)&={B}_{0}.
\end{aligned}\right.\end{aligned}
\end{equation}
This model is also called the reduced ECOM.
\par Another widely-used model is the ECOM2. This model was developed to describe the SRP of the GLONASS satellites. The definition of DYB frame here is the same as Eq.~(4). The ECOM2 model can be written as \cite{arnold_2015}
\begin{equation}
\begin{aligned}\left\{\begin{aligned}
    {D}(u)&={D}_{0}+\sum_{i=1}^{n_{D}}[{D}_{2i,c}\cos{2i\Delta{u}}+{D}_{2i,s}\sin{2i\Delta{u}}],\\
    {Y}(u)&={Y}_{0},\\
    {B}(u)&={B}_{0}+\sum_{i=1}^{n_{B}}[{B}_{2i-1,c}\cos{(2i-1)\Delta{u}}+{B}_{2i-1,s}\sin{(2i-1)\Delta{u}}].
\end{aligned}\right.\end{aligned}
\end{equation}
where $\Delta{u}=u-u_{s}$, and $u_s$ is the orbital latitude of the Sun.

\subsection{SRP models in the Cislunar Space}
\par In cislunar space, the orbital frequencies include the Earth-Moon barycenter's revolution around the Sun, the Moon's revolution around the Earth, and the satellite's revolution around the Earth. The current study focuses on orbits near the Lagrangian points and Moon-centered orbits (in synodic frame) such as DROs and NRHOs. Notably, a common characteristic of these orbit types is that these satellites revolve around a reference point, either the Lagrangian point or the Moon. As a preliminary analysis, the Sun, Earth, and Moon are assumed to lie in the same orbit plane. 
\begin{figure}[htp]
    \centering
    \subfigure[The planar Sun-Earth-Moon-Satellite geometry]{\includegraphics[width=7.5cm]{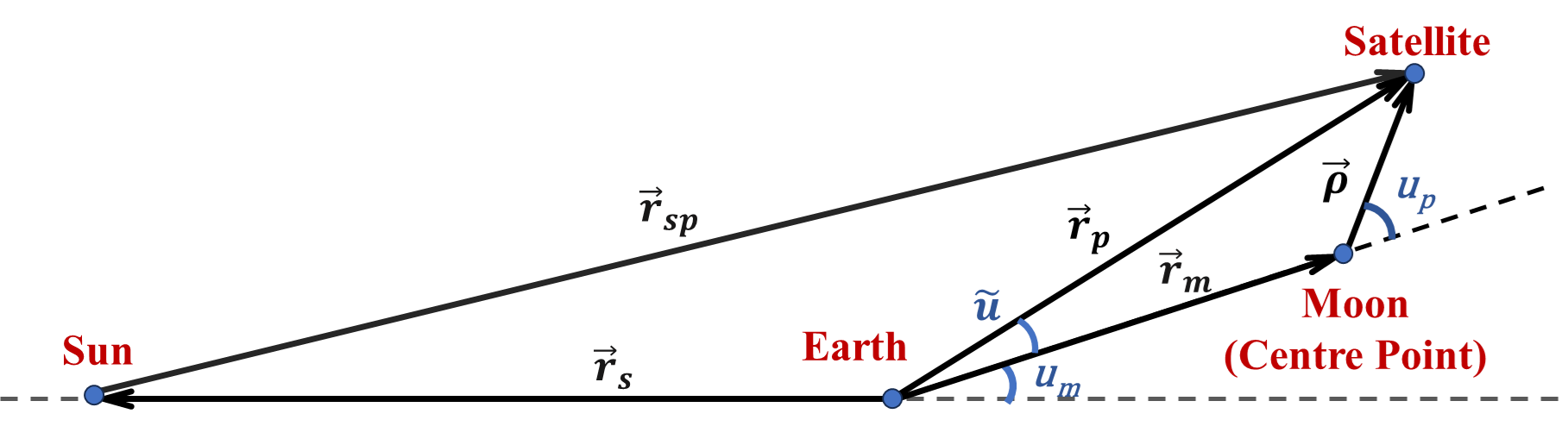}}
    \subfigure[The planar Sun-Earth-Satellite geometry]{\includegraphics[width=7.5cm]{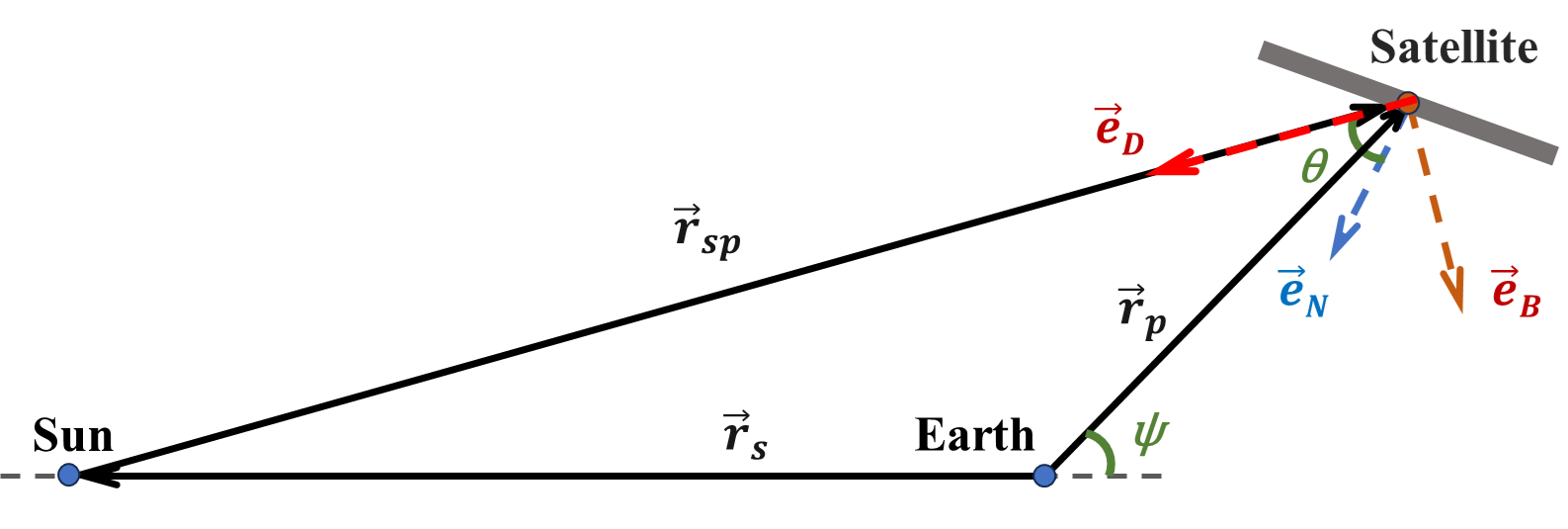}}
    \caption{Geometric relationships and parameter definitions}
    \label{fig:model}
\end{figure}

\par Firstly, several parameters are defined, as shown in Fig 3:
 \begin{itemize}
    \item $\mathbf{r}_s$ is the vector from the Earth to the Sun;
    \item $\mathbf{r}_m$ is the vector from the Earth to the Moon;
    \item $\mathbf{r}_p$ is the vector from the Earth to the satellite;
    \item $\mathbf{r}_{sp}$ is the vector from the Sun to the satellite;
    \item $\mathbf{e}_N$ is the normal direction of the surface;
    \item $\mathbf{e}_r$ is the normal direction of vector $\mathbf{r}_p$
    \item $\mathbf{e}_D$, $\mathbf{e}_Y$ and $\mathbf{e}_B$ are the unit vectors of the DYB frame defined in Eq.~(4);
    \item $\bm{\rho}$ is the vector from the Moon (or center point) to the satellite; 
    \item $u_m$ is the angle between $\mathbf{r}_{s}$ and $\mathbf{r}_{m}$
    \item $u_p$ is the angle between $\mathbf{r}_{m}$ and $\bm{\rho}$;
    \item $\widetilde{u}$ is the angle between $\mathbf{r}_p$ and $\mathbf{r}_m$;
    \item $\theta$ is the angle between $\mathbf{e}_D$ and $\mathbf{e}_N$
    \item $\psi=\widetilde{u}+u_m$ is the angle between $\mathbf{e}_D$ and $\mathbf{e}_r$
\end{itemize}
Typically, $r_s \gg r_m > \rho$. Thus, 
\begin{equation}
    {\gamma}_{1}={r}_{m}/{r}_{s} \ll 1 , \qquad {\gamma}_{2}=\rho/{r}_{s} \ll 1. 
\end{equation}
Based on the DYB frame defined by Eq.~(4), SRP can be expressed as 

\begin{equation}
\begin{aligned}
    &\mathbf{F}_{SRP}={F}_{D}\cdot\mathbf{e}_{D}+{F}_{Y}\cdot\mathbf{e}_{Y}+{F}_{B}\cdot\mathbf{e}_{B},\\
    &\left\{
    \begin{aligned}
    &{F}_{D}=-{F}_{1}-{F}_{2}\cos{\theta}=-\frac{{F}_{0}}{{r_{sp}}^{2}}\cos{\theta}[{P}_{1}+({P}_{2}+{P}_{3}\cos{\theta})\cos{\theta}],\\
    &{F}_{Y}=0,\\
    &{F}_{B}=-{F}_{2}\sin{\theta}=-\frac{{F}_{0}}{{r_{sp}}^{2}}\cos{\theta}\sin{\theta}({P}_{2}+{P}_{3}\cos{\theta}).
    \end{aligned}
    \right.
\end{aligned}
\end{equation}
where
\begin{equation}
    {F}_{0}=\frac{{P}_{0}S}{mc};\qquad
    {F}_{1}=\frac{{F}_{0}}{{r_{sp}}^{2}}\cos{\theta}{P}_{1};\qquad
    {F}_{2}=\frac{{F}_{0}}{{r_{sp}}^{2}}\cos{\theta}({P}_{2}+{P}_{3}\cos{\theta}).
    \nonumber
\end{equation}
\par In Eq.~(10), the angle $\theta$ is restricted to be in the range of $[-\pi/2, \pi/2]$, i.e., we require the normal direction of the plane point towards the Sun. It should be noted that there are two normal vectors for one plane. To obtain a continuous representation of $\theta$ over the full range $[0, 2\pi)$, a specific normal vector is fixed as $\mathbf{e}_N$. This choice allows one to determine which side of the surface faces the Sun based on the value of $\theta$. As shown in Fig. 4a, when $\theta \in [\pi/2, 3\pi/2)$, the back side of the surface faces the Sun. In this configuration, the vector $\mathbf{e}_D$ is opposite to $\mathbf{e}_N$. To account for both cases: $\theta \in [\pi/2, 3\pi/2)$ and $\theta \in [0, \pi/2) \cup [3\pi/2, 2\pi)$, the trigonometric functions in the final SRP expression require appropriate modification, as:
\begin{equation}
\begin{aligned}
    &\cos{\theta} \rightarrow (-1)^{[\frac{\theta+\pi/2}{\pi}]}\cos{\theta}=|\cos{\theta}|\\
    &\sin{\theta} \rightarrow (-1)^{[\frac{\theta+\pi/2}{\pi}]}\sin{\theta}\\
    &\cos{\theta}\sin{\theta} \rightarrow (-1)^{[\frac{\theta+\pi/2}{\pi}]}|\cos{\theta}|\sin{\theta}=\frac{1}{2}\sin{2\theta},
    \nonumber
\end{aligned}
\end{equation}
where $[ \cdot ]$ means the floor function.
\begin{figure}[htbp]
    \centering
    \subfigure[Geometrical display of the case when $\theta \in [\pi/2, 3\pi/2)$]{
        \includegraphics[width=7.5cm]{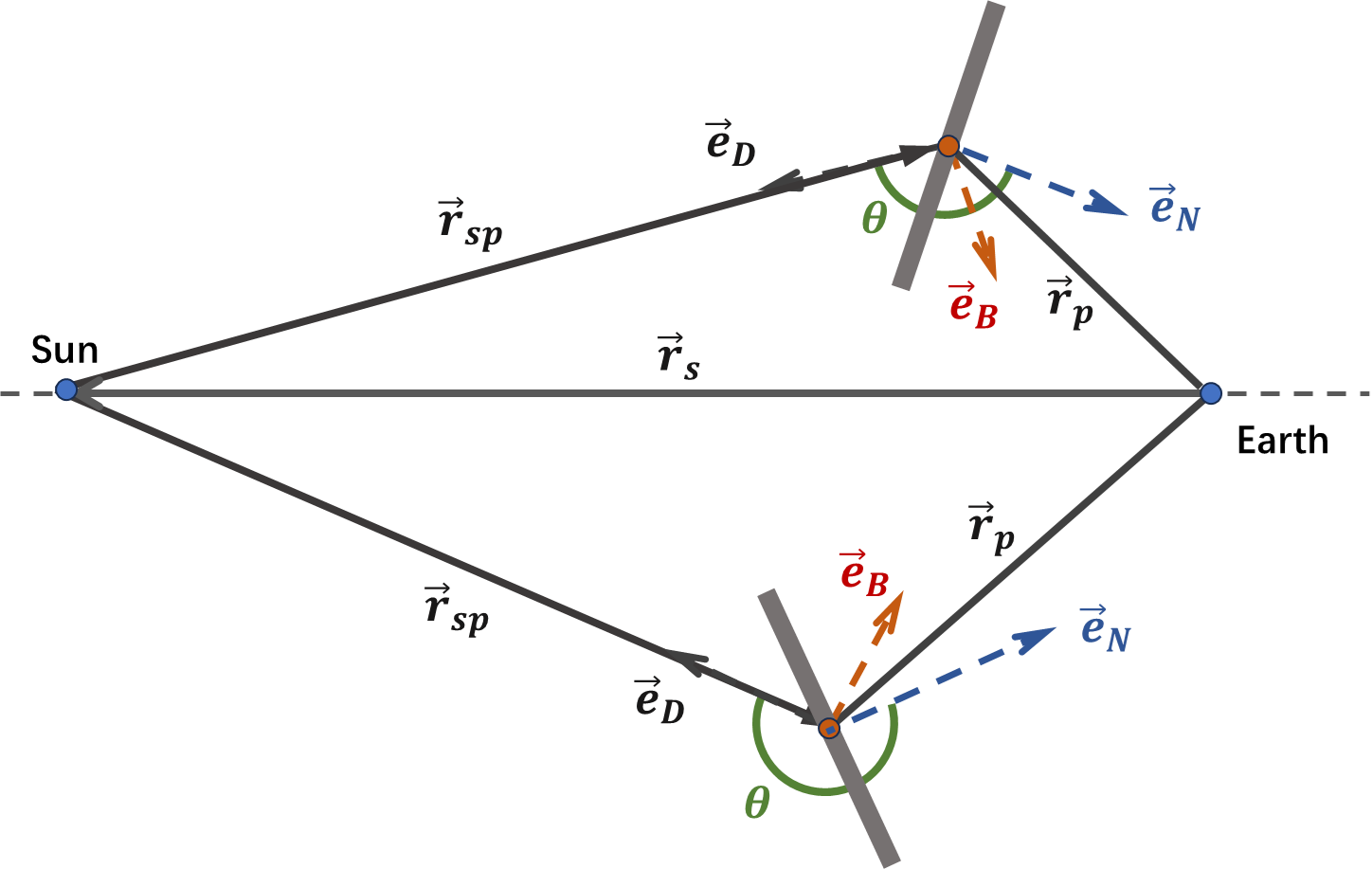}
        \label{fig:sub1}
    }
    \hfill
    \subfigure[Illustration of the abrupt reverse change when $\psi$ changes from smaller to larger than $\pi$]{
        \includegraphics[width=7.5cm]{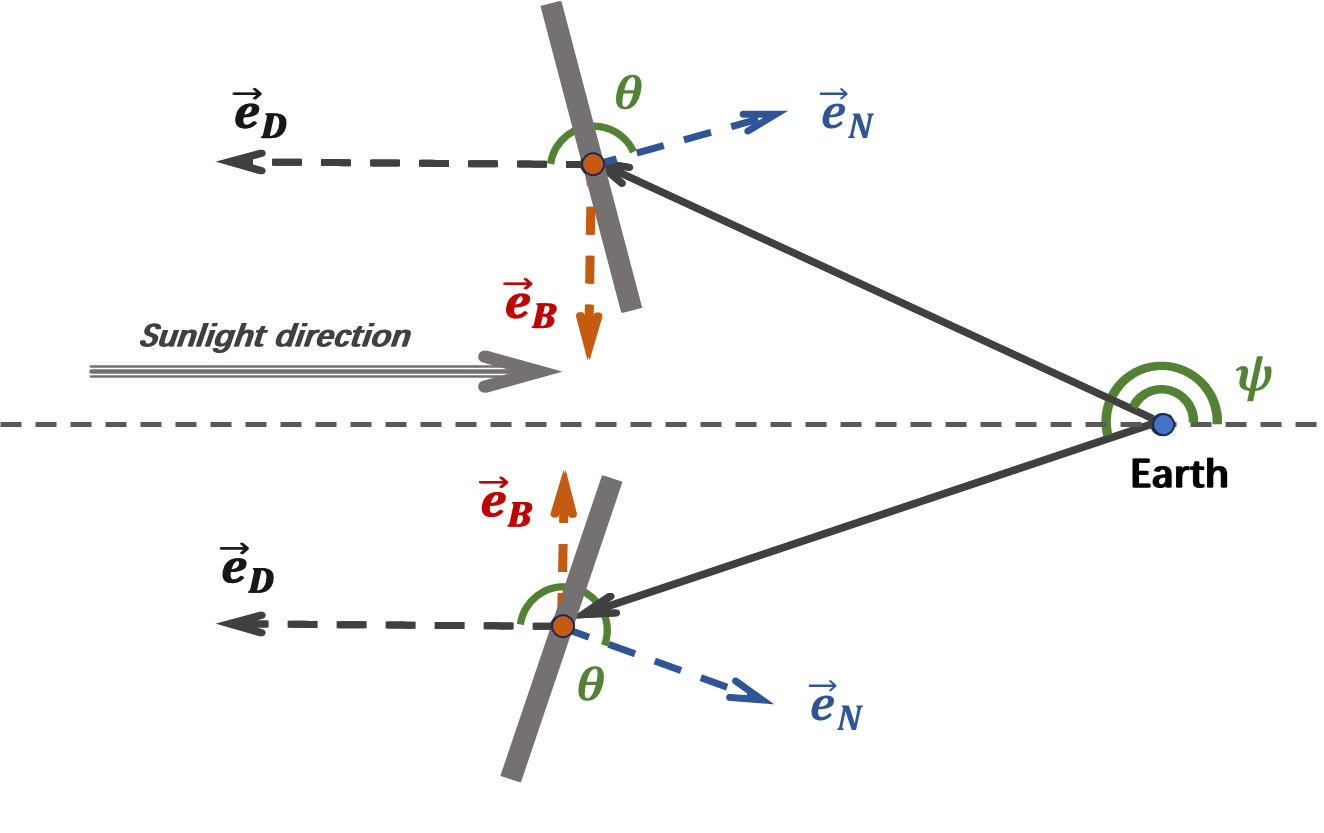}
        \label{fig:sub2}
    }
    \caption{Two geometric configurations}
    \label{fig:model}
\end{figure}
\par First, the SRP component in the $\mathbf{D}$ direction is analyzed. On account of $r_s\gg r_p$, the $\mathbf{D}$ direction can be considered approximately invariant and aligned with the direction from the Earth to the Sun, as shown in Fig. 4.b. Thus, the $\mathbf{D}$ direction force can be expressed as:
\begin{equation}
    {F}_{D}=-{F}_{1}-{F}_{2}\cos{\theta}
    =-\frac{{F}_{0}}{{r_{sp}}^{2}}|\cos{\theta}|({P}_{1}+{P}_{2}|\cos{\theta}|+{P}_{3}|\cos{\theta}|^{2}).\nonumber
\end{equation}
\par Next, the analysis proceeds to the $\mathbf{B}$ direction, where the scenario is more complex. Upon examining Fig. 4.b and comparing the cases where $0\leq\psi\leq\pi$ with those where $\psi\in[\pi,2\pi)$, it is apparent that, according to the definition in Eq.~(4), the $\mathbf{Y}$ and $\mathbf{B}$ directions undergo an abrupt reversal as $\psi$ crosses $\pi$. Considering this abrupt reversal in the $\mathbf{B}$ direction definition, ${F}_{B}$ should be expressed as:
\begin{equation}
    {F}_{B} = (-1)^{[\frac{\psi}{\pi}]}F_{2}\sin{\theta}
    = (-1)^{[\frac{\psi}{\pi}]}\frac{1}{2}\frac{{F}_{0}}{{r_{sp}}^{2}}\cos{2\theta}({P}_{2}+{P}_{3}|\cos{\theta}|).
    \nonumber
\end{equation}
\par In the expression above, the SRP component $F_B$ in the $\mathbf{B}$ direction depends on two angles: $\psi$ and $\theta$. The angle $\psi$ is related to the satellite's orbital motion, and the angle $\theta$ is related to the satellite's attitude, causing the frequency of $F_B$ to be more disordered in the current DYB frame. To accommodate this problem, a new direction vector $\mathbf{e}_M$ is introduced which is defined as the main orientation direction of the whole satellite, and the DYB coordinate can be redefined as:
\begin{equation}
    \mathbf{e}_{D}=\frac{\mathbf{r}_{sp}}{|{r}_{sp}|} \qquad
    \mathbf{e}_{Y}=\frac{\mathbf{e}_{D}\times\mathbf{e}_{M}}{|\mathbf{e}_{D}\times\mathbf{e}_{M}|} \qquad
    \mathbf{e}_{B}=\frac{\mathbf{e}_{D}\times\mathbf{e}_{Y}}{|\mathbf{e}_{D}\times\mathbf{e}_{Y}|}.
\end{equation}
Within this revised DYB frame, $\mathbf{e}_r$ is replaced by $\mathbf{e}_M$. Therefore, the angle $\psi$ needs to be redefined as the angle between $\mathbf{e}_D$ and $\mathbf{e}_M$. Consequently, when the satellite's attitude is fixed to a reference point (Sun, Earth, Moon, libration point, etc. depending on the different mission scenarios), the angle $\psi$ only differs from the angle $\theta$ by a constant angle, which is only determined by the orbital motion. This ensures that the frequencies of the two angles $\theta$ and $\psi$ are identical. Considering the fact that there are many faces of a satellite, and all these faces are generally fixed in the satellite's body-fixed frame, all these faces' angles (normal direction w.r.t. the $\mathbf{e}_D$ direction) are functions of the angle $\theta$. For a planar surface model, the main orientation coincides with the normal direction, i.e., $\mathbf{e}_M = \mathbf{e}_N$.Thus, SRP in the $\mathbf{D}$ and the $\mathbf{B}$ directions can be formulated as follows:
\begin{equation}
\begin{aligned}\left\{\begin{aligned}
    {F}_{D}&=-\frac{{F}_{0}}{{r_{sp}}^{2}}|\cos{\theta}|(P_{1}+P_{2}|\cos{\theta}|+P_{3}|\cos{\theta}|^{2}),\\
    {F}_{B}&=-(-1)^{[\frac{\theta}{\pi}]}\frac{{F}_{0}}{2{r_{sp}}^{2}}\sin{2\theta}({P}_{2}+{P}_{3}|\cos{\theta}|),
\end{aligned}\right.\end{aligned}
\end{equation}
where ${r}_{sp}$ is
\begin{equation}
\begin{aligned}
    r_{sp}&=\sqrt{[r_{s}+r_{m}\cos{u_{m}}+\rho\cos{(u_m+u_p)}]^2+[r_{m}\sin{u_{m}}+\rho\sin{(u_m+u_p)}]^2}\\
    &\approx {r}_{s}[1+{\gamma}_{1}\cos{u_{m}}+{\gamma}_{2}\cos{(u_m+u_p)}],\nonumber
\end{aligned}
\end{equation}
As a result, $1/r_{sp}^2$ can be expanded as follows:
\begin{equation}
\begin{aligned}    
    \frac{1}{r_{sp}^2}&={[(r_{s}+r_{m}\cos{u_{m}}+\rho\cos{(u_m+u_p)})^2+(r_{m}\sin{u_{m}}+\rho\sin{(u_m+u_p)})^2]}^{-\frac{1}{2}}\\
    &\approx \frac{1}{{r}_{s}^2}[1-2{\gamma}_{1}\cos{u_{m}}-2{\gamma}_{2}\cos{(u_m+u_p)}].
    \nonumber
\end{aligned}
\end{equation}
Similarly, other coefficients in Eq.~(12) can be expanded as Fourier series:
\begin{equation}
\begin{aligned}\left\{\begin{aligned}
    &|\cos{\theta}|= \sum_{i=0}^{\infty}C_{2i}\cos{2i\theta} = C_0+C_2\cos{2\theta} +C_4\cos{4\theta}+...\\
    &|\cos{\theta}|^{2} = |\cos^{2}{\theta}| = \cos^{2}{\theta} = \frac{1}{2}(\cos{2\theta}+1), \\
    &(-1)^{[\frac{\theta}{\pi}]} = \sum_{i=1}^{\infty}\frac{4}{(2n-1)\pi}\sin{2i\theta} = A_1\sin{\theta} +A_3\sin{3\theta}+...\nonumber
\end{aligned}\right.\end{aligned}
\end{equation}
\par Up to now, the analysis is restricted to the planar case, i.e., the satellite's orbit plane coincides with the Moon's orbit plane, and the SRP is parametrized by the angles $\theta$, $u_m$, and $u_p$. The following section demonstrates that the same analytical framework applies to the spatial (non-planar) case. However, the definitions of the angles $u_m$ and $u_p$ require modifications. As shown in Fig. 5, $\mathbf{e}_s$ is the direction from the Earth to the Sun; $\mathbf{e}_m$ is the direction from the Earth to the Moon; $\mathbf{e}_m'$ is the projection vector of $\mathbf{e}_m$ onto the satellite orbital plane in the Earth-Moon synodic frame. The angle $u_m$ retains its previous definition as the angle between $\mathbf{e}_s$ and $\mathbf{e}_m$. In contrast, $u_p$ is redefined as the angle between $\bm{\rho}$ and the projected vector $\mathbf{e}_m'$. 
\begin{figure}[htp]
    \centering
    \includegraphics[width=10cm]{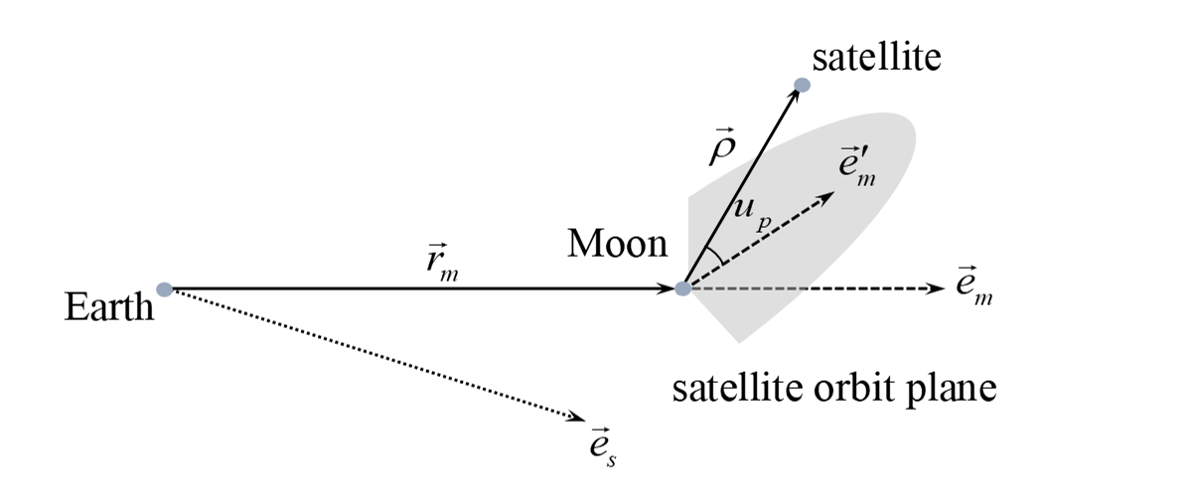}
    \caption{Schematic diagram of the spatial configuration of the Sun-Earth-Moon-Satellite geometry.}
\end{figure}
\par In the following subsections, we focus on some specific scenarios and expand $\theta$ as a series of expansions in terms of $u_m$ and $u_p$.

\subsubsection{Scenario 1: N direction fixed to the Sun}
\par In this scenario, the angle $\theta$ is constant. Consequently, Eq.~(12) can be rewritten as:
\begin{equation}
\begin{aligned}\left\{\begin{aligned}
    {F}_{D}&=-\frac{{F}_{0}}{{r_{sp}}^{2}}|\cos{\theta}|(P_{1}+P_{2}|\cos{\theta}|+P_{3}|\cos{\theta}|^{2})\\
    &= {K}_{sD}(1-2{\gamma}_{1}\cos{u_{m}}-2{\gamma}_{2}\cos{(u_m+u_p)}),\\
    {F}_{B}&=-(-1)^{[\frac{\theta}{\pi}]}\frac{{F}_{0}}{2{r_{sp}}^{2}}\sin{2\theta}({P}_{2}+{P}_{3}|\cos{\theta}|)\\
    &={K}_{sB}(1-2{\gamma}_{1}\cos{u_{m}}-2{\gamma}_{2}\cos{(u_m+u_p)}).
\end{aligned}\right.\end{aligned}
\end{equation}
Here
\begin{equation}
    {K}_{sD} = -\frac{{F}_{0}(P_{1}+P_{2}|\cos{\theta}|+P_{3}|\cos{\theta}|^{2})}{{r}_{s}^{2}}
    \quad and \quad 
    {K}_{sB} = -\frac{{F}_{0}\sin{2\theta}({P}_{2}+{P}_{3}|\cos{\theta}|)}{2{r}_{s}^{2}},
    \nonumber
\end{equation}
are constants. In this case, the SRP variation depends only on the distance from the Sun to the satellite $r_{sp}$. Usually, the angle $\theta$ is small enough that $F_B$ can be considered as a constant (similar to the reduced ECOM). Under this approximation, SRP can be simplified as:
\begin{equation}
\begin{aligned}\left\{\begin{aligned}
    {F}_{D}&= {K}_{sD}(1-2{\gamma}_{1}\cos{u_{m}}-2{\gamma}_{2}\cos{(u_m+u_p)}),\\
    {F}_{B}&= {K}_{sB}.
\end{aligned}\right.\end{aligned}
\end{equation}

\subsubsection{Scenario 2: N direction fixed to the Earth}
\par In this scenario, the N direction is aligned with the direction from the satellite to the Earth, and the angle $\theta\approx\psi$. Accordingly, $F_D$ can be expanded as:
\begin{equation}
\begin{aligned}
    {F}_{D}&=-\frac{{F}_{0}}{{r_{sp}}^{2}}|\cos{\theta}|(P_{1}+P_{2}|\cos{\theta}|+P_{3}|\cos{\theta}|^{2})\\
    &\approx -\frac{{F}_{0}}{{r}_{s}^{2}}[1-2{\gamma}_{1}\cos{u_{m}}-2{\gamma}_{2}\cos{(u_m+u_p)}](D_0+D_2\cos{2\theta}+D_4\cos{4\theta}).
\end{aligned}
\end{equation}
Similarly, the expansion of $F_B$ up to the third order is given by:
\begin{equation}
\begin{aligned}
    {F}_{B}&=-(-1)^{[\frac{\theta}{\pi}]}\frac{{F}_{0}}{2{R}^{2}}\sin{2\theta}({P}_{2}+{P}_{3}|\cos{\theta}|)\\
    &\approx -\frac{{F}_{0}}{2{r}_{s}^{2}}[1-2{\gamma}_{1}\cos{u_{m}}-2{\gamma}_{2}\cos{(u_m+u_p)}](B_1\cos{\theta}+B_3\cos{3\theta}).
\end{aligned}
\end{equation}
Here $D_0$, $D_2$, $D_4$, $B_1$, and $B_3$ are coefficients. Given that ${\gamma}_{1}={r}_{m}/{r}_{s} \ll 1 $ and ${\gamma}_{2}=\rho/{r}_{s} \ll 1$, the primary variation of the SRP is governed by the angle $\theta$. We first expand $1/r_p$ as:
\begin{equation}
\begin{aligned}
    \frac{1}{r_p}&=\frac{1}{\sqrt{(r_m)^2+\rho^2+2r_m\rho\cos{u_p}}}
    = \frac{1}{r_m}\frac{1}{\sqrt{1-2\kappa(-\cos{u_p})+\kappa^2}}\\
    &= \frac{1}{r_m}\sum^\infty_{m=0}P_m(-\cos{u_p})\kappa^m
    = \frac{1}{r_m}\left[(1-\frac{1}{2}\kappa^2)-\kappa\cos{u_p}+\frac{3}{2}\kappa^2\cos^2{u_p}\right],
    \nonumber
\end{aligned}
\end{equation}
where $\kappa=\rho/r_m$. The Legendre polynomials $P_m(x)$ are
\begin{equation}
    P_0(x) = 1, \qquad P_1(x) = x, \qquad P_2(x) = \frac{3}{2}x^2-\frac{1}{2}. \nonumber
\end{equation}
The expression is truncated to second order in $\kappa$, and the same truncation applies to subsequent expansions. Consequently, $\widetilde{u}$ can be derived as: 
\begin{equation}
\begin{aligned}
    \cos{\widetilde{u}}&=\frac{r_p^2+r_m^2-\rho^2}{2rr_m} = 1-\frac{\kappa^2}{4}+\frac{\kappa^2}{4}\cos{2u_p}\\
    \sin{\widetilde{u}}&=\frac{\rho}{r_p}\sin{u_p} = \kappa\sin{u_p}-\frac{\kappa^2}{2}\sin{2u_p}
    \nonumber
\end{aligned}.
\end{equation}
Thus, the angle $\theta$ satisfies the following relationship:
\begin{equation}
\begin{aligned}
    \cos{\theta}&= \cos{(u_m+\widetilde{u})}=\cos{u_m}\cos{\widetilde{u}}-\sin{u_m}\sin{\widetilde{u}}\\
    &= \left(1-\frac{\kappa^2}{4}\right)\cos{u_m}-\frac{\kappa}{2}\cos{(u_p-u_m)}+\frac{\kappa}{2}\cos{(u_p+u_m)}. \nonumber
\end{aligned}
\end{equation}
Substituting the above relations into Eq.~(12), the SRP can be expressed as:
\begin{equation}
\begin{aligned}\left\{\begin{aligned}
    F_{D} &= -\frac{F_{0}}{R^{2}}\Bigl\{
        [D_0 + D_2\cos 2u_m + D_4\cos 4u_m] 
        + \kappa D_2[\cos(u_p + 2u_m) - \cos(u_p - 2u_m)] \\
    &\quad + 2\kappa D_4[\cos(u_p + 4u_m) - \cos(u_p - 4u_m)]
        \Bigr\}, \\[6pt]
F_{B} &= -\frac{F_{0}}{R^{2}}\Bigl\{
        (B_1\cos u_m + B_3\cos 3u_m)
        + \frac{\kappa B_1}{2}\bigl[\cos(u_p + u_m) - \cos(u_p - u_m)\bigr] \\
    &\quad + \frac{3\kappa B_3}{2}\bigl[\cos(u_p + 3u_m) - \cos(u_p - 3u_m)\bigr]
        \Bigr\}.
\end{aligned}\right.\end{aligned}
\end{equation}
\par When $\kappa$ is small, terms involving $\kappa$ can be neglected, and the expressions simplify to functions of $u_m$ only:
\begin{equation}
\begin{aligned}\left\{\begin{aligned}
    {F}_{D}&=-\frac{{F}_{0}}{{r_s}^{2}}(D_0+D_2\cos{2u_m}+D_4\cos{4u_m}),\\
    {F}_{B}&= -\frac{{F}_{0}}{2{r_s}^{2}}(B_1\cos{u_m}+B_3\cos{3u_m}).
\end{aligned}\right.\end{aligned}
\end{equation}

\subsubsection{Scenario 3: N direction fixed to the Moon or centre point}
\par In this scenario, the N direction aligns with the direction from the satellite to the geometrical centre of its orbit  (e.g., the Moon for DROs/NRHOs, or the triangular libration point for TLPOs). First, the planar orbit case (such as the DRO) is considered, where the satellite, the Earth, and the Moon are approximately coplanar. Therefore, $\theta = u_m + u_p$ since $\mathbf{r}_{sp} \approx -\mathbf{r}_s$. After some derivations, the SRP can be derived as
\begin{equation}
\begin{aligned}\left\{\begin{aligned}
    {F}_{D}&= -\frac{{F}_{0}}{{r_s}^{2}}[({P}_{1}{C}_{0}+\frac{P_{2}}{2}+\frac{P_{3}C_0}{2}+\frac{P_{3}C_2}{4})\\
    &\qquad +({P}_{1}{C}_{2}+\frac{P_{2}}{2}+P_{3}\frac{C_0+C_2}{2})\cos{2(u_m + u_p)}+\frac{P_{3}C_2}{4}\cos{4(u_m + u_p)}],\\
    {F}_{B}&= -\frac{{F}_{0}}{2{r_s}^{2}}[\left[\frac{A_1+A_3}{2}(P_2+\frac{P_3C_0}{2})+\frac{A_1P_3C_2}{8}\right]\cos{(u_m + u_p)}\\
    &\quad\qquad -\left(\frac{A_1}{2}(P_2+\frac{P_3C_0}{2})+\frac{A_3P_3C_2}{8}\right)\cos{3(u_m + u_p)}].
\end{aligned}\right.\end{aligned}
\end{equation}
\par For the NRHO, the orbit is nearly perpendicular to the Earth-Moon plane. In this case, $\cos{\theta}=\cos{u_p}\cdot\cos{u_m}$. Using this relation, the SRP for the perpendicular case is:
\begin{equation}
\begin{aligned}\left\{\begin{aligned}
    {F}_{D}&= -\frac{F_0}{8r_s^2}[(2D_0-4D_2+D_4)+2D_2\cos (2u_p+2u_m)+2D_2\cos (2u_p-2u_m)\\
    &\qquad +(-4D_2+4D_4)\cos2u_p+(-4D_2+4D_4)\cos 2u_m+3D_4\cos4u_p+3D_4\cos4u_m],\\
    {F}_{B}&= -\frac{F_0}{16r_s^2}[(4B_1-3B_3) \cos(u_p-u_m)+(-4B_1+3B_3) \cos(u_p+u_m) \\
    &\qquad -3B_3 \cos (3u_p-u_m)+3B_3\cos (3u_p+u_m)+3B_3\cos (u_p+3 u_m)-3B_3\cos (u_p-3u_m)].
\end{aligned}\right.\end{aligned}
\end{equation}

\subsubsection{Scenarios 4$\&$5: N direction facing a fixed point in the Earth-Moon system}
\par A more general scenario is now considered. As shown in Fig. 6(a), the N direction points towards a fixed reference point. The angle $\psi'$ is defined as the satellite-fixed point-centre point angle; the angle $\sigma$ is the fixed point-centre point-Earth angle; $L$ is the distance from the satellite to the fixed point; $r_f$ is the distance from the centre point to the fixed point. From geometry, $\theta=\psi'+u_m-\sigma$, and $\sigma$ is constant. Since a constant phase shift does not affect the frequency content of the SRP, the specific value of $\sigma$ can be disregarded. In the following derivation, the angle $\sigma$ is set as 0 for simplicity, as illustrated in Fig. 6(b).
\begin{figure}[htp]
    \centering
    \subfigure[]{\includegraphics[width=7.5cm]{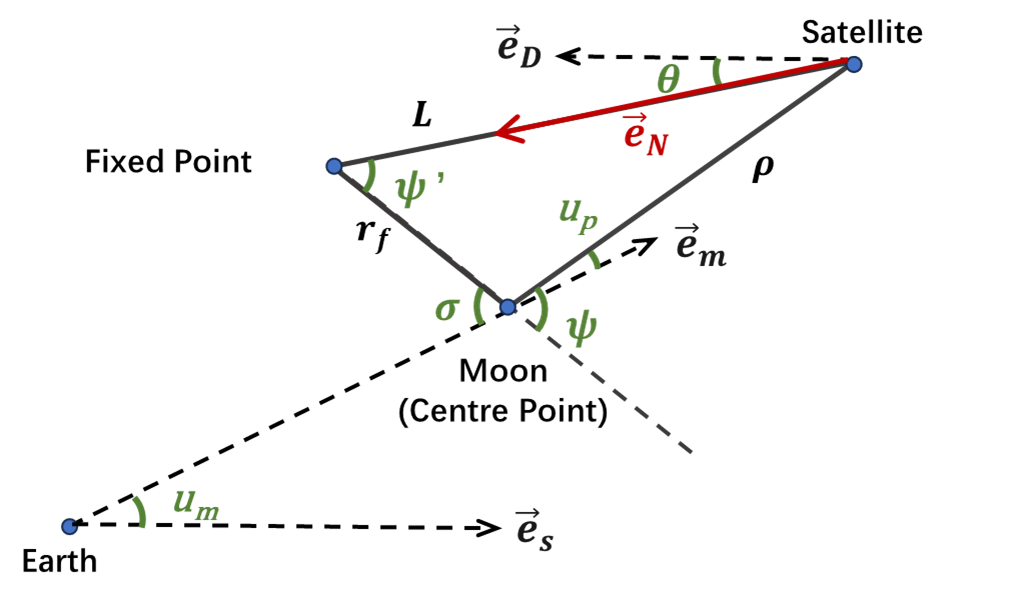}}
    \subfigure[]{\includegraphics[width=7.5cm]{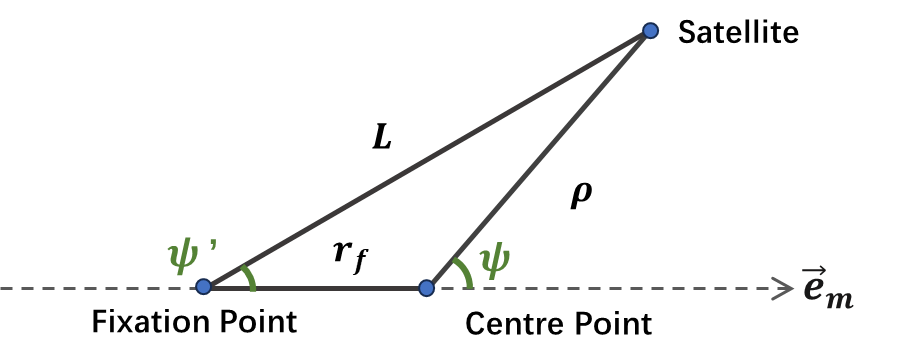}}
    \caption{Geometry among fixed point, Earth and satellite}
    \label{fig:model}
\end{figure}

\par Similar to the scenarios above, the angles $\psi'$ and $\theta$ need to be expanded. This case is separated into two distinct scenarios: $r_f<\rho$ (Scenario 4) and $r_f>\rho$ (Scenario 5). In Scenario 4, a small parameter $k$ is defined as $k = r_f / \rho$. In Scenario 5, $k$ is defined as $k = \rho / r_f$. The same expansion method as in Scenario 2 is applied. Thus, for Scenario 4, the angle $\theta$ can be derived as
\begin{equation}
\begin{aligned}
    \cos{\theta}&= \cos{(u_m+\psi')}= \cos{u_m}\cos{\psi'}-\sin{u_m}\sin{\psi'}\\
    &= \frac{k}{2}\cos{u_m}-\frac{1}{8}k^2\cos{(u_p-u_m)}+\left(1-\frac{1}{4}k^2\right)\cos{(u_p+u_m)}\\
    &\quad-\frac{k}{2}\cos{(2u_p+u_m)}+\frac{3k^2}{8}\cos{(3u_p+u_m)}.\nonumber
\end{aligned}
\end{equation} 
Substituting this relation into Eq.~(12), the SRP for Scenario 4 is expressed as:
\begin{equation}
\begin{aligned}\left\{\begin{aligned}
    {F}_{D}&= -\frac{F_0}{r_{s}^2}\{D_0+(1-k^2)D_2\cos{(2u_p+2u_m)}+(1-4k^2)D_4\cos{(4u_p+4u_m)}\\
    &\quad +kD_2[\cos{(u_p+2u_m)}+\cos{(3u_p+2u_m)}]+2kD_4[\cos{(3u_p+4u_m)}-\cos{(5u_p+4u_m)}]\\
    &\quad +{k^2}{D_2}\cos{(4u_p+2u_m)}+{k^2}{D_4}[\cos{(2u_p+4u_m)}+3\cos{(6u_p+4u_m)}]\},\\
    {F}_{B}&= -\frac{F_0}{2r_{s}^2}\{B_1\left(1-\frac{k^2}{4}\right)\cos{(u_p+u_m)}+B_3\left(1-\frac{9k^2}{4}\right)\cos{(3u_p+3u_m)}\\
    &\quad +\frac{B_1k}{2}[\cos{u_m}+\cos{(2u_p+u_m)}]+\frac{3B_3k}{2}[\cos{(2u_p+3u_m)}+\cos{(4u_p+3u_m)}]\\ &\quad +\frac{B_1k^2}{8}[-\cos{(u_p-u_m)}+3\cos{(3u_p+u_m)}]+\frac{3B_3k^2}{8}[\cos{(u_p+3u_m)}+5\cos{(5u_p+3u_m)}]\}.
\end{aligned}\right.\end{aligned}
\end{equation}
\par Analogously, similar expressions for Scenario 5 can be obtained:
\begin{equation}
\begin{aligned}\left\{\begin{aligned}
    {F}_{D}&= -\frac{F_0}{R^2}\{D_0+(1-k^2)D_2\cos{2u_m}+(1-4k^2)D_4\cos{4u_m}\\
    &\quad +kD_2[\cos{(u_p+2u_m)}-\cos{(u_p-2u_m)}]+2kD_4[\cos{(u_p+4u_m)}-\cos{(u_p-4u_m)}]\\
    &\quad +{k^2}{D_2}\cos{(2u_p-2u_m)}+{k^2}{D_4}[3\cos{(2u_p-4u_m)}+\cos{(2u_p+4u_m)}]\},\\
    {F}_{B}&= -\frac{F_0}{2R^2}\{B_1\left(1-\frac{k^2}{4}\right)\cos{u_m}+B_3\left(1-\frac{9k^2}{4}\right)\cos{3u_m}\\
    &\quad +\frac{B_1k}{2}[\cos{(u_p+u_m)}-\cos{(u_p-u_m)}]+\frac{3B_3k}{2}[\cos{(u_p+3u_m)}-\cos{(u_p-3u_m)}]\\ &\quad +\frac{B_1k^2}{8}[3\cos{(2u_p-u_m)}-\cos{(2u_p+u_m)}]+\frac{3B_3k^2}{8}[\cos{(2u_p+3u_m)}+5\cos{(2u_p-3u_m)}]\}.
\end{aligned}\right.\end{aligned}
\end{equation}
\par Evidently, as $k \rightarrow 0$, the expression for Scenario 4 reduces to that of Scenario 3, where the N direction is fixed to the Moon or the centre point. Similarly, the expression for Scenario 5 reduces to that of Scenario 2, where the N direction is fixed to the Earth.

\section{SRP recovery analysis}
\par The preceding section proposed new SRP models for different scenarios, expressed as combinations of the angles $u_p$ and $u_m$. However, these derivations assumed an initial phase angle $\beta = 0$ between $\bm{\rho}$ and $\mathbf{e}_m$. In a more realistic setting, this phase difference is non-zero and must be accounted for. Therefore, each trigonometric term of the form $\cos(\alpha)$ in the previous expressions should be generalized to:
\begin{equation}
    C\cos{(\alpha+\beta)}=C_1\cos{\alpha}+C_2\sin{\alpha}.
    \nonumber
\end{equation}
Consequently, modeling each angular combination requires two coefficients. However, an excessive number of parameters can degrade the OD accuracy. Thus, the strategy adopted here is to select a limited set of key frequencies and evaluate the accuracy of the SRP model constructed from them.
\par Following the theoretical framework from Section 2.1, this section employs the box-wing model to generate reference SRP data (simulating the real SRP). For the selected frequencies, the corresponding model coefficients are then estimated from this simulated data via a least-square fit. The accuracy of the new parametric SRP model is subsequently assessed by comparing the box-wing model output against the fitted results from these models. This comparison validates the precision and reliability of the proposed modeling approach. All initial orbital parameters used in the simulations are provided in Appendix A.

\subsection{Establishment of SRP simulation}
\par In this study, the box-wing model is employed to simulate the real SRP. The satellite is decomposed into 10 planar surfaces, comprising a main body (box) and two solar panels (wings). Specific values are assigned to the satellite mass $m$, the area $S$ of each surface, and its optical coefficients ($\alpha$, $\rho$, $\delta$). These values are listed in Table 1. Using this geometric model together with the satellite’s attitude information, the illuminated faces can be determined and the simulated SRP can be computed according to Eq.~(5). 
\begin{figure}[htp]
    \centering
    \includegraphics[width=7.5cm]{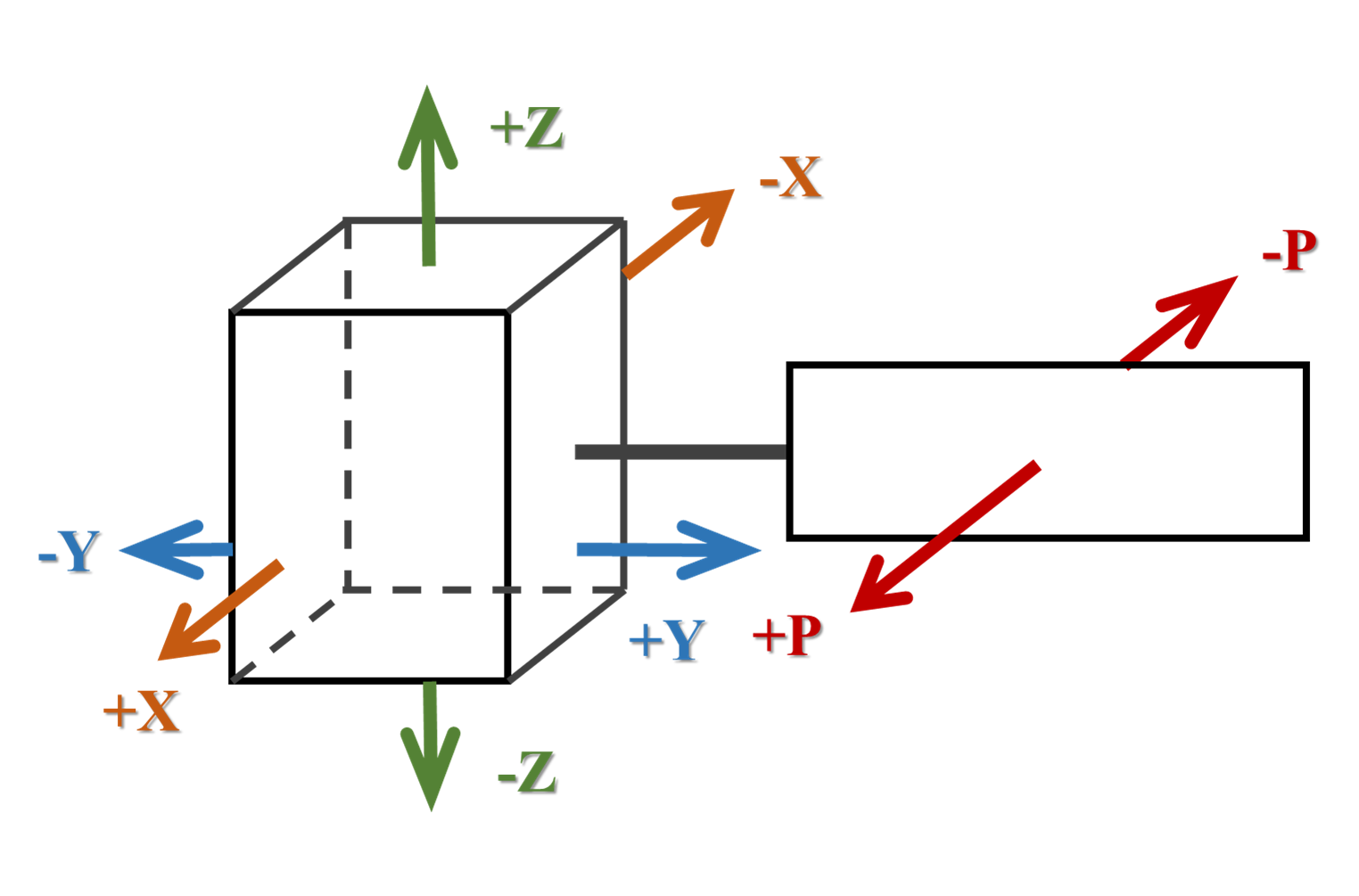}
    \caption{Schematic of the satellite box-wing model.}
\end{figure}
\par The outward normal directions of the 10 surfaces are set as follows. As shown in Fig. 7, the box possesses six normal directions labelled as $+\mathbf{X}$, $-\mathbf{X}$, $+\mathbf{Y}$, $-\mathbf{Y}$, $+\mathbf{Z}$, $-\mathbf{Z}$, while the wings have two opposite normal directions: $+\mathbf{P}$, $-\mathbf{P}$. The $+\mathbf{X}$ direction is set as the main orientation axis of the satellite $+\mathbf{e}_M$. A surface is considered non‑illuminated when the angle between the normal direction and the Sun direction is larger than $\pi$. Mutual shadowing between surfaces is neglected in these simulations.
\begin{equation}
\begin{aligned}
    \mathbf{e}_{+X}=\mathbf{e}_{M};\quad \mathbf{e}_{-X}=-\mathbf{e}_{M};\quad &\mathbf{e}_{+P}=\mathbf{e}_{M};\quad \mathbf{e}_{-P}=-\mathbf{e}_{M};\quad\\
    \mathbf{e}_{+Y}=\frac{\mathbf{e}_{M}\times\mathbf{e}_{rc}}{|\mathbf{e}_{M}\times\mathbf{e}_{rc}|};\quad \mathbf{e}_{-Y}=-\mathbf{e}_{+Y}; \quad
    &\mathbf{e}_{+Z}=\frac{\mathbf{e}_{+X}\times\mathbf{e}_{+Y}}{|\mathbf{e}_{+X}\times\mathbf{e}_{+Y}|};\quad \mathbf{e}_{-Z}=-\mathbf{e}_{+Z}.\\
    \nonumber
\end{aligned}
\end{equation}
This setting yields the fact that there is almost no SRP component in the $Y$ direction. Situations with non-zero $F_Y$ are treated separately in Section~5. In this paper, all of our simulations adopt the following settings: the mass of the satellite $m$ is set as 1000kg; the area of each plane is shown in Table~\ref{Surface_set} and optical coefficients of all surface are set to $\alpha=0.3$, $\rho=0.5$, $\delta=0.2$.

\begin{table}[htp]
    \renewcommand\arraystretch{1.2}
    \centering
    \begin{threeparttable} 
    \caption{\textbf{Surface areas in the box‑wing model used for the simulations.}}
    \begin{tabular}{|c c|c c|}
    \hline
    Surface & Area/$m^2$ & Surface & Area/$m^2$ \\ \hline
    Wing Front/Back & 150 & Box $+\mathbf{Y}/-\mathbf{Y}$ & 50  \\ \hline
    Box $+\mathbf{X}/-\mathbf{X}$ & 50 & Box $+\mathbf{Z}/-\mathbf{Z}$ & 20 \\ \hline
    \end{tabular}
    \label{Surface_set}
    \end{threeparttable} 
\end{table}

\subsection{The general new SRP model}

\par For practical application, a general SRP model that comprehensively encompasses all scenarios outlined in Section~2.3 is desirable. Neglecting angular combinations with small amplitudes, the final expression of the general model is given by Eq.~(23). This model is termed the Empirical NJU Cislunar Model (\textbf{ENCM}).

\begin{equation}
\begin{aligned}\left\{\begin{aligned}
    {F}_{D}&= D_1+D_2\cos{2u_m}+D_3\sin{2u_m}+D_4\cos{2u_p}+D_5\sin{2u_p}\\
    & \quad +D_6\cos{2(u_p+u_m)}+D_7\sin{2(u_p+u_m)}+D_8\cos{2(u_p-u_m)}\\
    & \quad +D_9\sin{2(u_p-u_m)}+D_{10}\cos{(u_p+2u_m)}+D_{11}\sin{(u_p+2u_m)}\\
    & \quad +D_{12}\cos{(3u_p+2u_m)}+D_{13}\sin{(3u_p+2u_m)}\\
    & \quad +D_{14}\cos{(u_p-2u_m)}+D_{15}\sin{(u_p-2u_m)},\\
    {F}_{Y}&= Y_1,\\
    {F}_{B}&= B_1+B_2\cos{u_m}+B_3\sin{u_m}+B_4\cos{3u_m}+B_5\sin{3u_m}\\
    & \quad +B_6\cos{(u_p+u_m)}+B_7\sin{(u_p+u_m)}+B_8\cos{3(u_p+u_m)}+B_9\sin{3(u_p+u_m)}\\
    & \quad +B_{10}\cos{5(u_p+u_m)}+B_{11}\sin{5(u_p+u_m)}+B_{12}\cos{(u_p-u_m)}\\
    & \quad +B_{13}\sin{(u_p-u_m)}+B_{14}\cos{(2u_p+u_m)}+B_{15}\sin{(2u_p+u_m)}.\\
\end{aligned}\right.\end{aligned}
\end{equation}

\par The validity of the general model is verified by fitting it, via the least‑square principle, to the SRP accelerations generated by the box‑wing model. Several representative test cases are presented here. Their configurations are listed in Table~\ref{tab:setting_of_tests}, and the fitting results are shown in Fig.~8. Initial orbital elements for these tests are provided in Appendix~A. Considering the characteristic periods of cislunar orbits, a 14‑day data arc is chosen for the SRP inversion. For Test~5, the fixed point is located on the Earth–Moon line at a distance of 0.05~$L_{\text{EM}}$ from the Moon, where~$L_{\text{EM}}$ denotes the mean Earth–Moon distance. The results clearly demonstrate that the ENCM can accurately reproduce the simulated SRP, justifying its use in the OD process.  

\begin{table}[htp]
    \renewcommand\arraystretch{1.2}
    \centering
    \begin{threeparttable} 
    \caption{\textbf{Settings of the SRP fitting tests.}}
    \begin{tabular}{c|c c c c}
    \hline
    Test ID & Orbit type & Scenario & Fixed point & Arc length (days) \\ \hline
    Test 1 & DRO Small & 1 & Sun & 14 \\ \hline
    Test 2 & DRO Large & 2 & Earth & 14 \\ \hline
    Test 3 & DRO Small & 3 & Moon & 14 \\ \hline
    Test 4 & NRHO North & 3 & Moon & 14 \\ \hline
    Test 5 & DRO Large & 4 & Point at 0.05~$L_{\text{EM}}$ \tnote{a} & 14 \\ \hline
    Test 6 & L4 Large & 5 & Moon & 14 \\ \hline
    \end{tabular}
    \begin{tablenotes}
        \small
        \item[a] Point on the Earth–Moon line, 0.05 times the Earth–Moon distance ($L_{\text{EM}}$) from the Moon.
    \end{tablenotes}
    \label{tab:setting_of_tests}
    \end{threeparttable} 
\end{table}

\begin{figure}[htp]
    \centering
    \subfigure[Test 1]{\includegraphics[width=4.5cm]{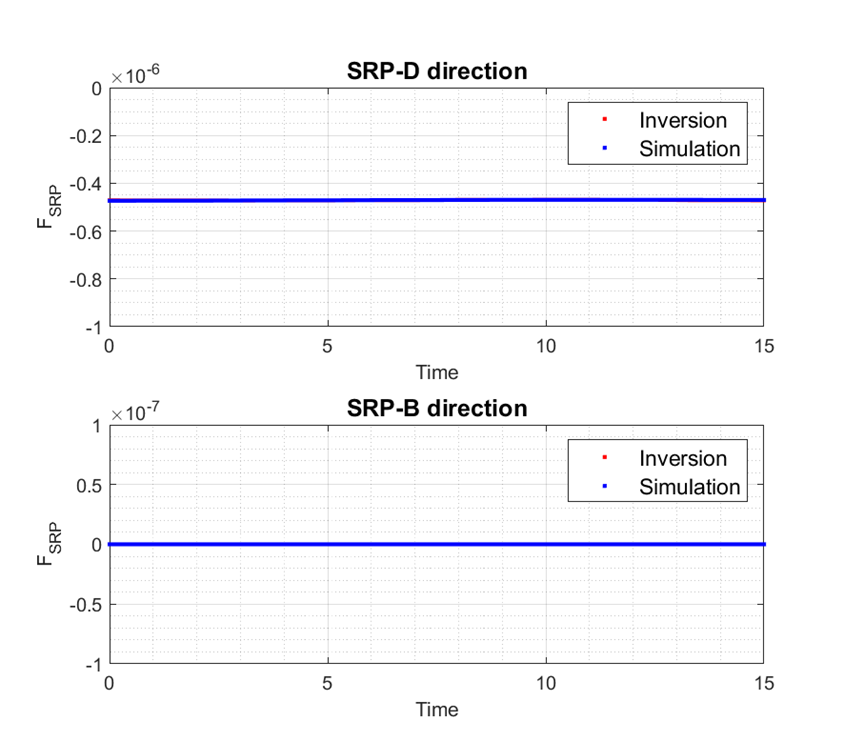}}
    \subfigure[Test 2]{\includegraphics[width=4.5cm]{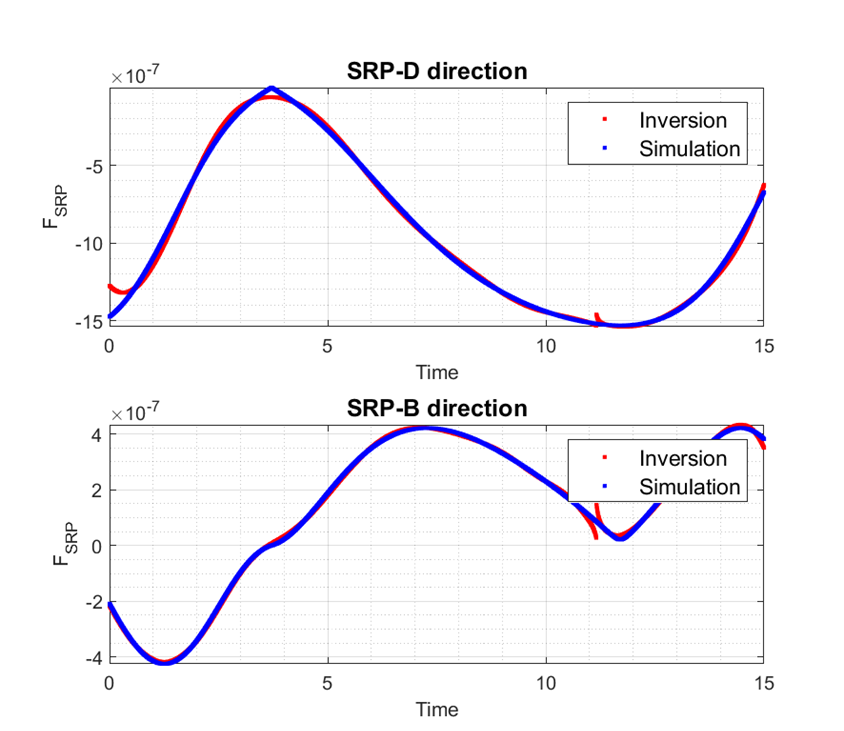}}
    \subfigure[Test 3]{\includegraphics[width=4.5cm]{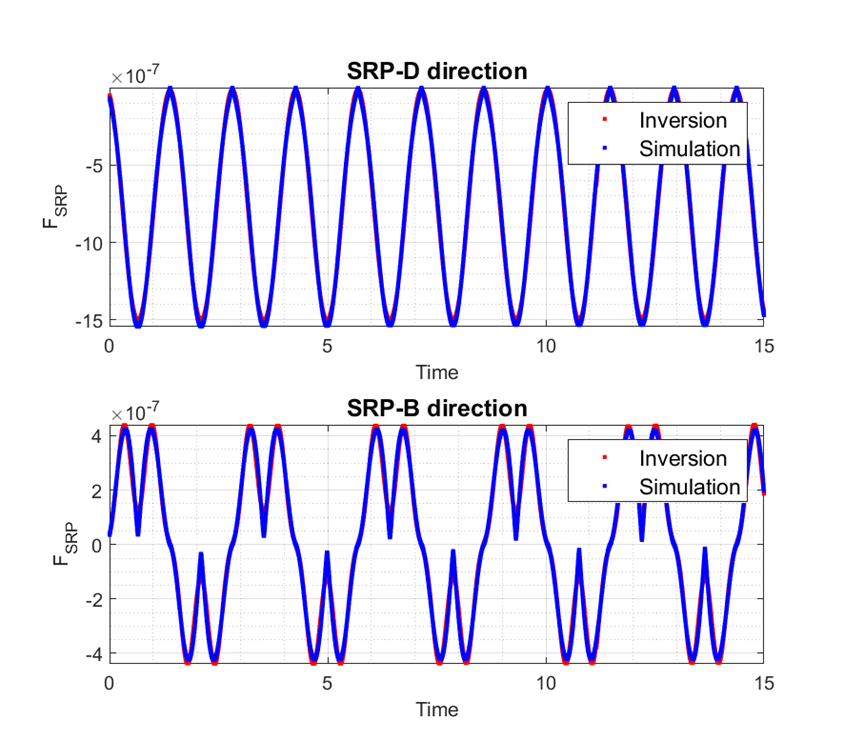}}
    \subfigure[Test 4]{\includegraphics[width=4.5cm]{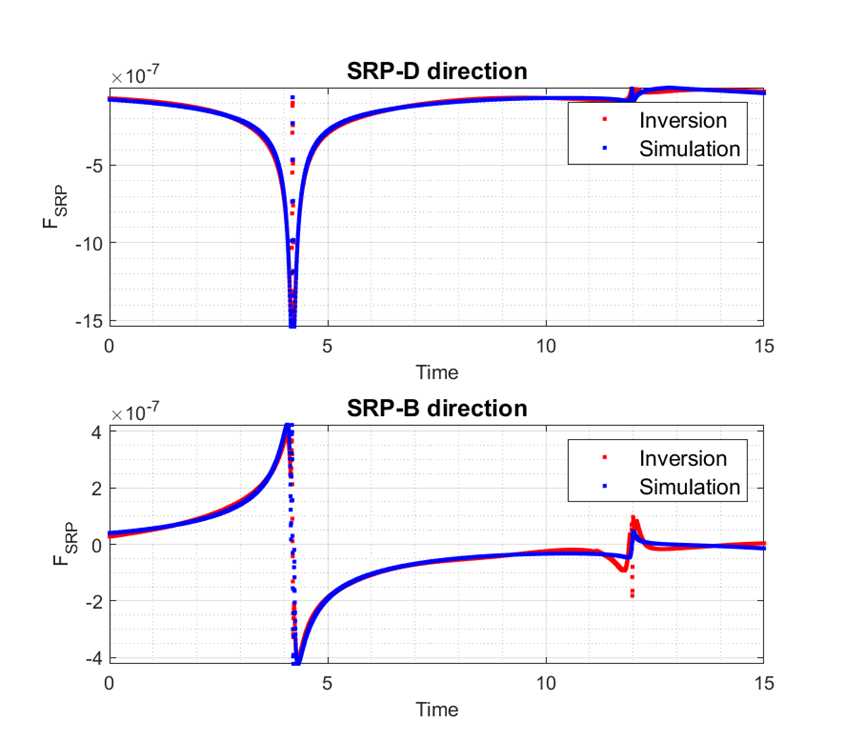}}
    \subfigure[Test 5]{\includegraphics[width=4.5cm]{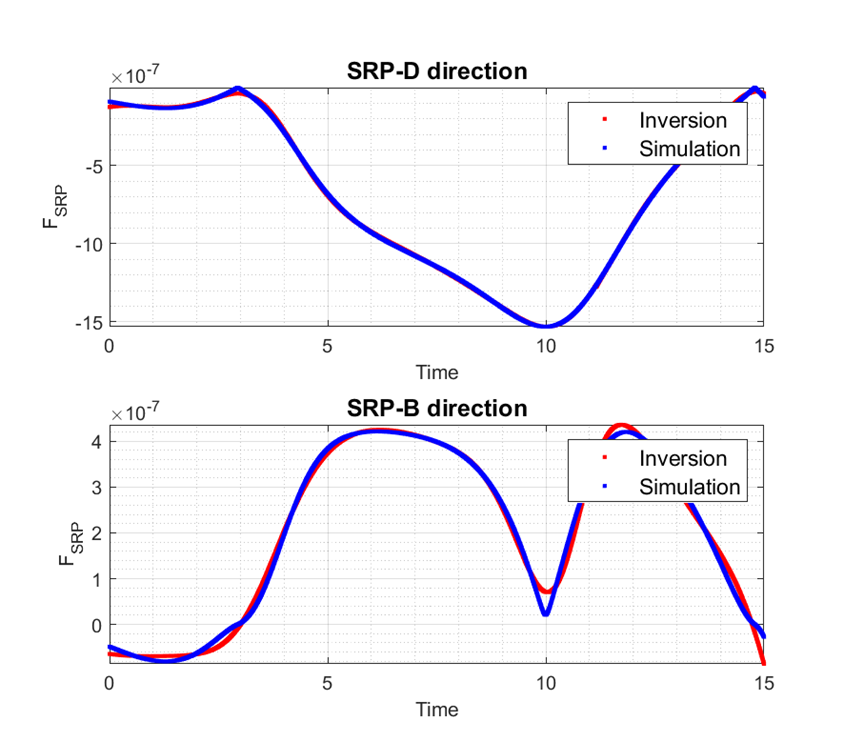}}
    \subfigure[Test 6]{\includegraphics[width=4.5cm]{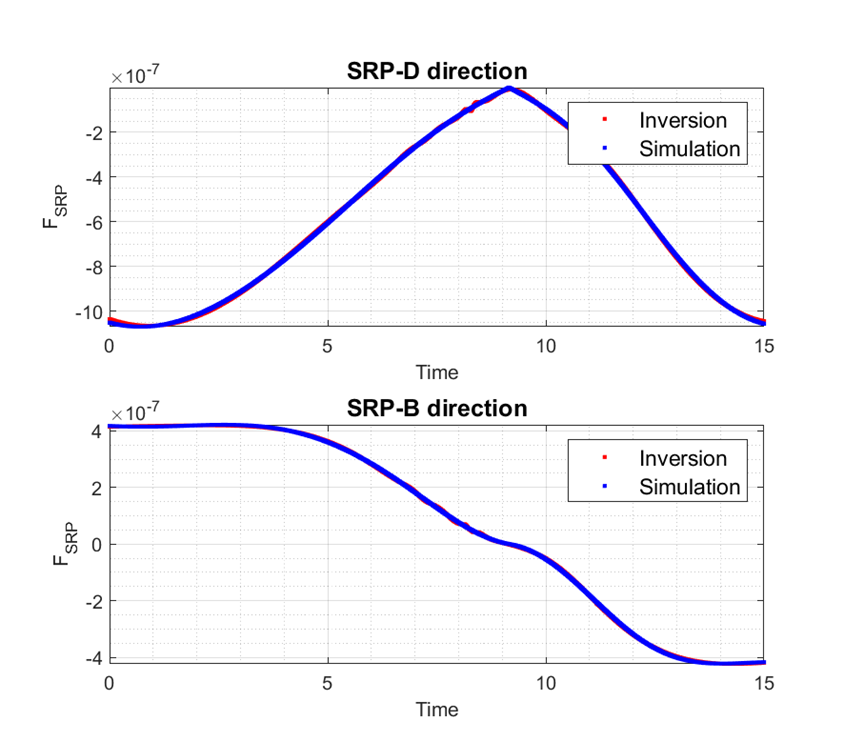}}
    \caption{Comparison of the simulated (blue) and fitted (red) SRP components (in $m/s^2$) using Eq.~(23).}
    \label{fig:scenarios}
\end{figure}

\newpage
\section{Application in the OD}
\par This section applies the ENCM within the OD process. The advantage of the ENCM is demonstrated by comparing the OD results against those obtained using either the cannonball model or the ECOM model. 

\subsection{The Batch Algorithm}
\par The initial state vector at the reference epoch is denoted as $\mathbf{X}_0$, the set of parameters to be estimated (here, the SRP coefficients) as $\mathbf{C}$, and the observation vector as $\mathbf{Y}$. For each observation $y_i$, the residual between the estimated and observed values is:
\begin{equation}
y_i=Y_{ei}-Y_{oi}=H_x\hat{\mathbf{x}}_{0}+H_c\hat{\mathbf{c}}+\epsilon,\nonumber
\end{equation}
where $Y_{oi}$ is the $i^{\text{th}}$ observed value, $Y_{ei}$ is the corresponding estimated value, $\hat{\mathbf{x}}_{0}$ is the correction to the current orbital estimate, $\hat{\mathbf{c}}$ is the correction to the current parameter estimate, and $\epsilon$ represents higher-order terms neglected in the OD process. The matrices $H_x$ and $H_c$ are defined as:
\begin{equation}
\begin{aligned}
H_x=\frac{\partial Y_{ei}}{\partial \mathbf{X_i}}\frac{\partial \mathbf{X_i}}{\partial \mathbf{X_0}}=\widetilde{H_x}\Phi;& \quad \Phi=\frac{\partial \mathbf{X_i}}{\partial \mathbf{X_0}};\quad \widetilde{H_x}=\frac{\partial Y_{ei}}{\partial \mathbf{X_i}},\\
H_c=\frac{\partial Y_{ei}}{\partial \mathbf{X_i}}\frac{\partial \mathbf{X_i}}{\partial \mathbf{C}}=\widetilde{H_x}\Phi_c;& \quad \Phi_c=\frac{\partial \mathbf{X_i}}{\partial \mathbf{C}}.
\nonumber
\end{aligned}
\end{equation}
Here, $\Phi$ is the state transition matrix (STM) and $\Phi_c$ is the parameter sensitivity matrix (PSM). Their time evolution is governed by the variational equations:
\begin{equation}
\begin{aligned}\left\{\begin{aligned}
& \Dot{\Phi}(\mathbf{X},t)=\frac{\partial \Dot{\mathbf{X}}}{\partial \mathbf{X}}\Phi(\mathbf{X},t),\\
& \Dot{\Phi}_c(\mathbf{X},t)=\frac{\partial \Dot{\mathbf{X}}}{\partial \mathbf{X}}\Phi_c(\mathbf{X},t)+\frac{\partial{ \Dot{\mathbf{X}}}}{\partial C}.
\end{aligned}\right.\end{aligned}
\end{equation}
Defining the combined design matrix $H = [H_x,H_c]$ and assuming equally weighted observations, the corrections $\hat{\mathbf{x}}_{0}$ and $\hat{\mathbf{c}}$ are obtained from the least-squares solution:
\begin{equation}
\left[\begin{array}{c} \hat{\mathbf{x}}_0 \\ \hat{\mathbf{c}} \end{array}\right]=\left[\sum^{n}_{i=1}(H^TH)\right]^{-1}\sum^{n}_{i=1}H^Ty_i.
\end{equation}
The OD process iterates until the corrections $\hat{\mathbf{x}}_{0}$ and $\hat{\mathbf{c}}$ become sufficiently small or a predefined maximum number of iterations is reached \cite{schutz_2004}.

\subsection{The OD results}
\par An in-house orbit determination (OD) software developed by the authors is used for the simulations. This software operates in the Barycentric Celestial Reference System (BCRS) using Barycentric Dynamical Time (TDB), although the initial conditions provided in Appendix~A are expressed in the Earth-centered celestial reference system. This software considers:
\begin{itemize}
    \item Earth's non-spherical gravity, expanded to degree and order~5 using the EGM~2008 model.
    \item Moon's non-spherical gravity, expanded to degree and order ~2 using the GRAIL-derived model GGGRX1200a.
    \item 3rd-body gravitational perturbations from the Sun, Moon, and planets. Positions are obtained from the JPL DE440 ephemeris \cite{schutz_2004}.
    \item Relativistic corrections.  With TDB as the time scale, these are provided by the Einstein-Infeld-Hoffmann (EIH) formulation \cite{soffel_2013}.
    \item Earth's tidal forces, modeled according to the IERS 2010 conventions \cite{petit_2010}.
    \item Coupling between Earth's non-spherical gravity and the gravitational fields of major bodies \cite{brumberg_2004}.
    \item Solar radiation pressure (SRP).
\end{itemize}

 \par The observable is the inter-satellite link (ISL) range. Three inclined geosynchronous orbit (IGSO) satellites, with precisely known orbits, serve as reference space platforms. The ISL links are simulated between these platforms and the cislunar satellites whose states are to be estimated by the OD process. The initial states of the three IGSO satellites (labeled IGSO-1, IGSO-2, and IGSO-3) are also listed in Appendix~A. The reference (“true”) SRP is still simulated using the box-wing model (see Section~3.1). The light-time delay between satellites is accounted for. To isolate the effect of SRP modeling error, no random measurement noise is added to the simulated ISL data.

\begin{table}[htp]
    \renewcommand\arraystretch{1.2}
    \centering
    \small
    \setlength{\tabcolsep}{4pt}  
    \begin{threeparttable}
    \caption{\textbf{Comparison of the OD accuracy using different SRP models.}}
    \begin{tabular}{|c|c|c|c|c|c|c|c|c|}
    \hline
    \multirow{2}{*}{ID} & \multirow{2}{*}{Orbit Type} & \multirow{2}{*}{Fixed Point}  & \multicolumn{2}{|c|}{Cannonball} & \multicolumn{2}{|c|}{ECOM}  & \multicolumn{2}{|c|}{ENCM} \\ \cline{4-9}
    & & & RMS (m) & Max (m) & RMS (m) & Max (m) & RMS (m) & Max (m) \\ \hline
    1 & DRO Small & the Sun & 0.01 & 0.02 & 0.06 & 0.20 & $\mathbf{1.70}$ & $\mathbf{4.67}$ \\  \hline
    2 & DRO Large & the Earth & 20655 & 49356 & 5144.69 & 9160.32 & $\mathbf{16.48}$ & $\mathbf{59.81}$ \\  \hline
    3 & NRHO North & the Earth & 1087.7 & 2393.6 & 272.2 & 1539.7 & $\mathbf{67.80}$ & $\mathbf{265.54}$ \\  \hline
    4 & L4 Orbit & the Earth & 31357 & 44700 & 8752.4 & 30707 & $\mathbf{198.58}$ & $\mathbf{449.28}$ \\  \hline
    5 & DRO Small & the Moon & 892.1 & 1947.3 & 340.7 & 669.2 & $\mathbf{7.67}$ & $\mathbf{1.79}$ \\  \hline
    6 & NRHO North & the Moon & 668.9 & 988.8 & 19.65 & 93.58 & $\mathbf{12.61}$ & $\mathbf{5.47}$ \\  \hline
    7 & DRO Large & Point at 0.05~$L_{\text{EM}}$ \tnote{a}  & 31551 & 41842 & 4587.6 & 8988.9 & $\mathbf{23.21}$ & $\mathbf{51.75}$ \\  \hline
    8 & L4 Orbit & the Moon & 17956 & 2004.3 & 4196.9 & 104374 & $\mathbf{17.34}$ & $\mathbf{65.11}$ \\  \hline
    \end{tabular}
    \begin{tablenotes}
        \footnotesize
        \item[a] Point on the Earth–Moon line, 0.05 times the Earth–Moon distance ($L_{\text{EM}}$) from the Moon.
    \end{tablenotes}
    \label{tab:od_results_comparison}
    \end{threeparttable} 
\end{table}

\par Table~\ref{tab:od_results_comparison} summarizes the OD accuracy for different orbit types and scenarios by using different SRP models. Test~1 corresponds to Scenario~1 in Section~2.3; Tests~2--4 correspond to Scenario~2; Tests~5 and~6 correspond to Scenario~3; Test~7 corresponds to Scenario~4; and Test~8 corresponds to Scenario~5. Except for Tests~3 and~6, which use a 4-day data arc, all other tests employ a 14-day arc. The rationale for the shorter arc for NRHO cases will be explained in Section~5. For a clear visual comparison, the root-mean-square (RMS) position errors of the estimated orbits relative to the true orbits are presented as a bar chart in Fig.~9, where the vertical axis uses a base-10 logarithmic scale.

\begin{figure}[htp]
    \centering
    \includegraphics[width=18cm]{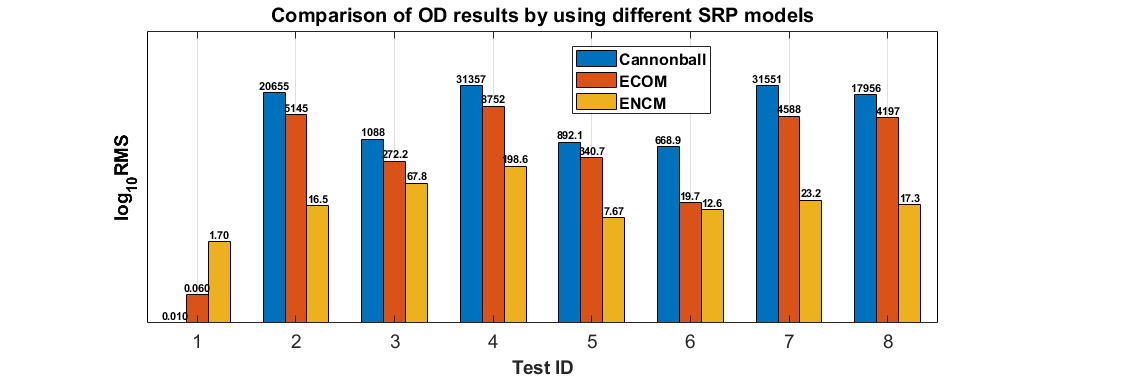}
    \caption{Comparison of OD results using different SRP models.}
\end{figure}

\par The results confirm that the ENCM performs well across all tested scenarios. It demonstrates superior accuracy in most cases, with the notable exception of Test~1, where the satellite's principal orientation is fixed toward the Sun. In that specific scenario, the cannonball model provides a perfect SRP representation, whereas both the ECOM and ENCM introduce additional parameters. In an ideal fit, these parameters should be zero; however, estimation errors prevent them from converging exactly to zero, introducing bias and resulting in slightly degraded performance compared to the cannonball model. Nevertheless, for all the other tests, the ENCM yields significantly better OD accuracy. Figs.~10 and~11 compare the SRP accelerations estimated using the ENCM against the reference SRP generated by the box-wing model. The estimated SRP closely matches the reference values, validating the ENCM's fidelity. These findings reinforce the potential of the ENCM for practical application in high-precision orbit determination.

\begin{figure}[htp]
    \centering
    \subfigure[OD position error.]{\includegraphics[width=4.5cm]{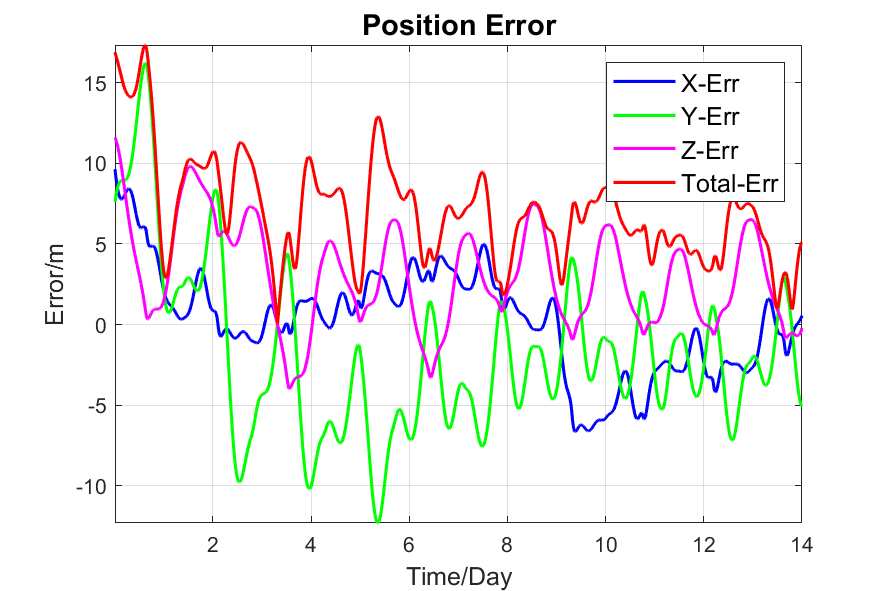}}
    \subfigure[SRP estimated with the ENCM.]{\includegraphics[width=10cm]{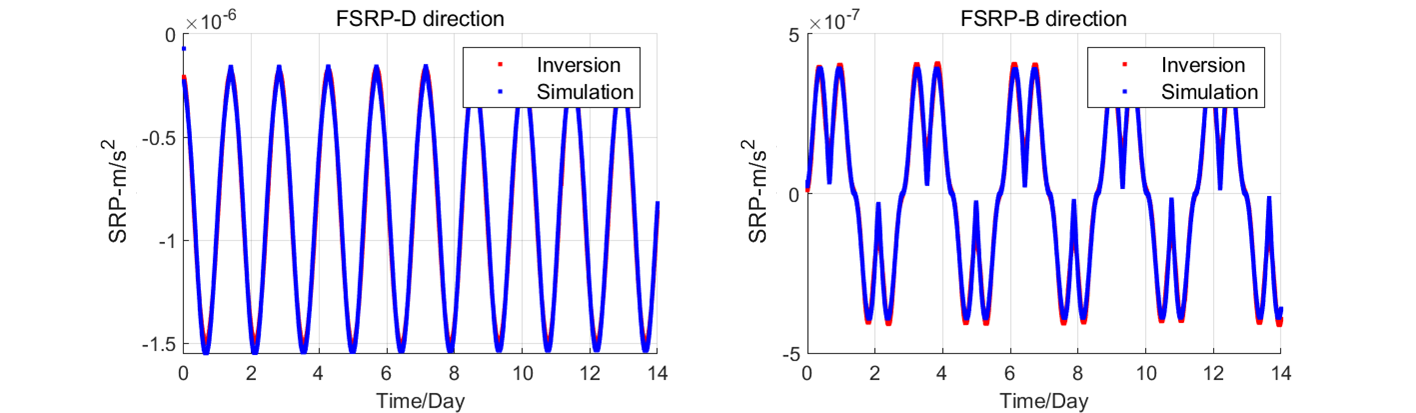}}
    \caption{OD results for Test 5 using the ENCM.}
    \label{fig:test5_results}
\end{figure}

\begin{figure}[htp]
    \centering
    \subfigure[OD position error.]{\includegraphics[width=4.5cm]{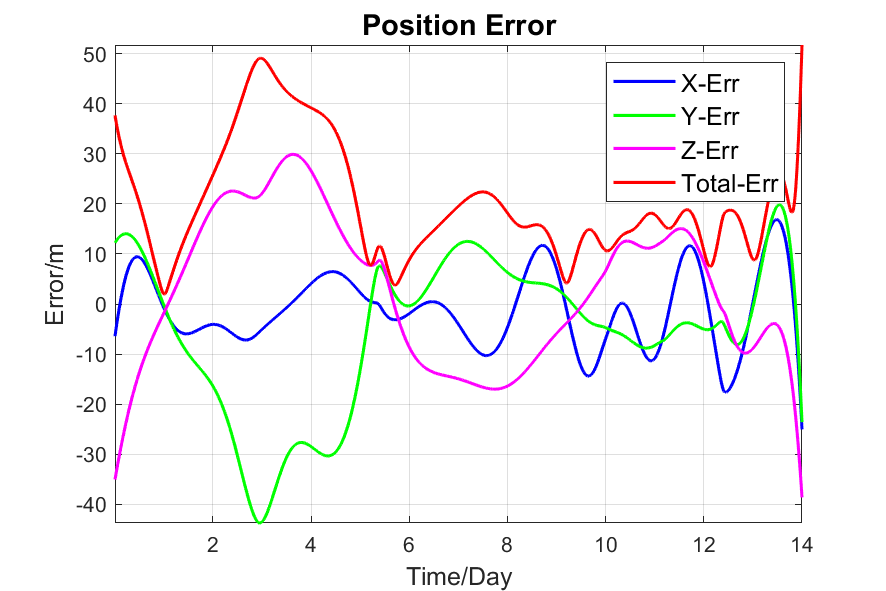}}
    \subfigure[SRP estimated with the ENCM.]{\includegraphics[width=10cm]{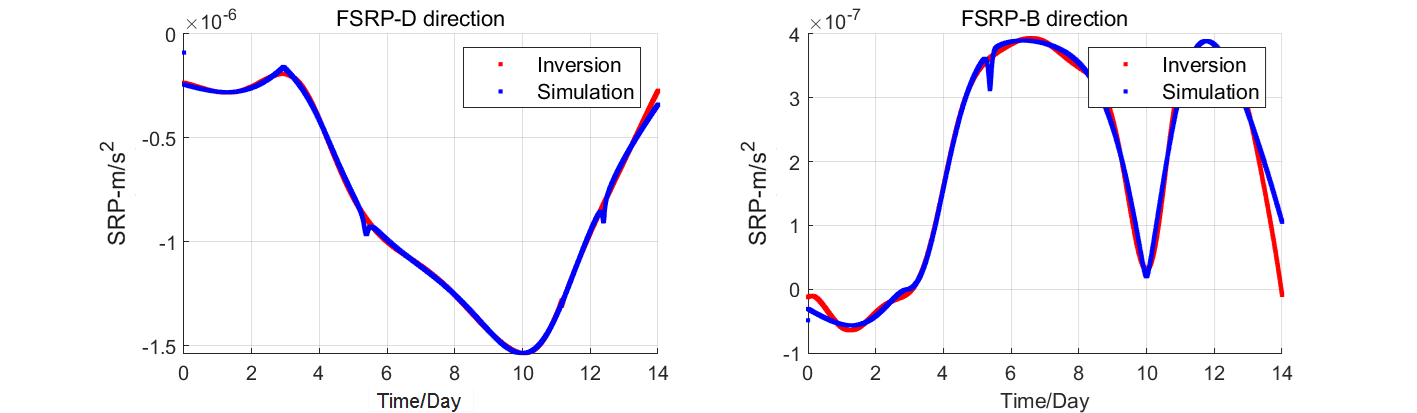}}
    \caption{OD results for Test 7 using the ENCM.}
    \label{fig:test7_results}
\end{figure}

\section{Discussion}
\subsection{Sub-models}

\par Eq.~(23) presents a general ENCM capable of accommodating all scenarios described in Section~2.3. However, its large number of parameters typically demands a long data arc and a high observation count for reliable estimation. In practice, for any given scenario, the coefficients of many angular combinations in the general model are intrinsically zero. Including these unnecessary terms in the force model and estimating them in the OD process often leads to non-zero estimated values. Moreover, the mis-estimation of these spurious terms can corrupt the estimation of the truly non-zero coefficients. Therefore, if the working scenario of the satellite is known a priori, it is advantageous to employ the corresponding sub-model (termed Sub-ENCM) specifically developed for that scenario. For Scenarios~2--5, the specific model forms derived from Eqs.~(18)--(22) are summarized, with their complete expressions provided in Appendix~B.

\begin{table}[htp]
    \renewcommand\arraystretch{1.2}
    \caption{\textbf{Comparison of the OD results using general ENCM and Sub-ENCM}}
    \centering
    \begin{tabular}{|c|c|c|c|c|c|c|}
    \hline
    \multirow{2}{*}{Arc length (days)} & \multirow{2}{*}{Orbit Type} & \multirow{2}{*}{Fixed Point}  & \multicolumn{2}{|c|}{General ENCM} & \multicolumn{2}{|c|}{Sub-ENCM}  \\ \cline{4-7}
    & & & RMS (m) & Max (m) & RMS (m) & Max (m) \\ \hline
    7 & DRO Small & Moon & 9.09 & 19.91 & 0.94 & 2.02 \\  \hline
    \end{tabular}
    \label{}
\end{table}

\par Due to more accurate modelling of the force model in these scenarios, the Sub-ENCMs can significantly improve the OD accuracy in cases of shorter arcs. An example is shown in Table~4 and Figs.~12--13. This test uses a 7-day arc and corresponds to Scenario~5. The results clearly show that the Sub-ENCMs achieve higher OD accuracy than the general ENCM.

\begin{figure}[htp]
    \centering
    \subfigure[OD position error.]{\includegraphics[width=4.5cm]{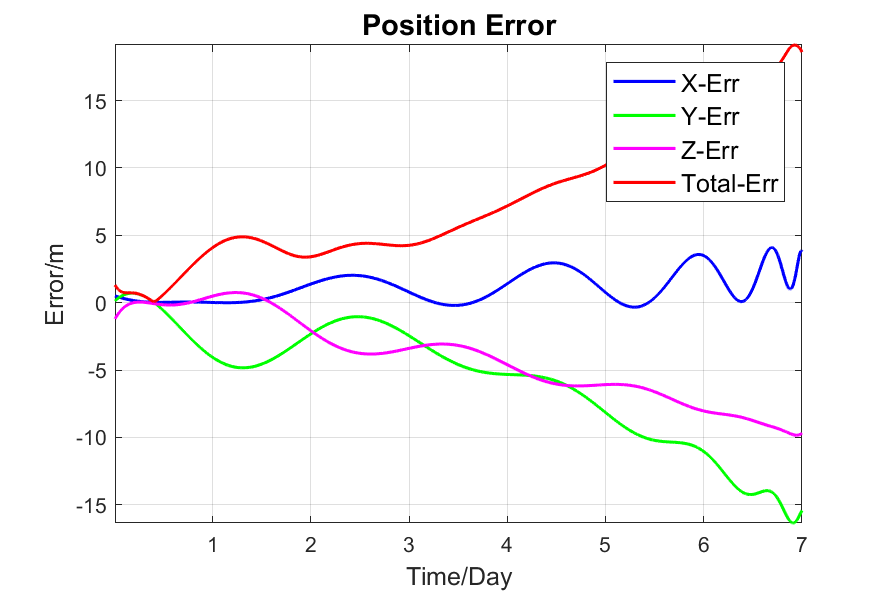}}
    \subfigure[SRP estimated with the general ENCM.]{\includegraphics[width=9cm]{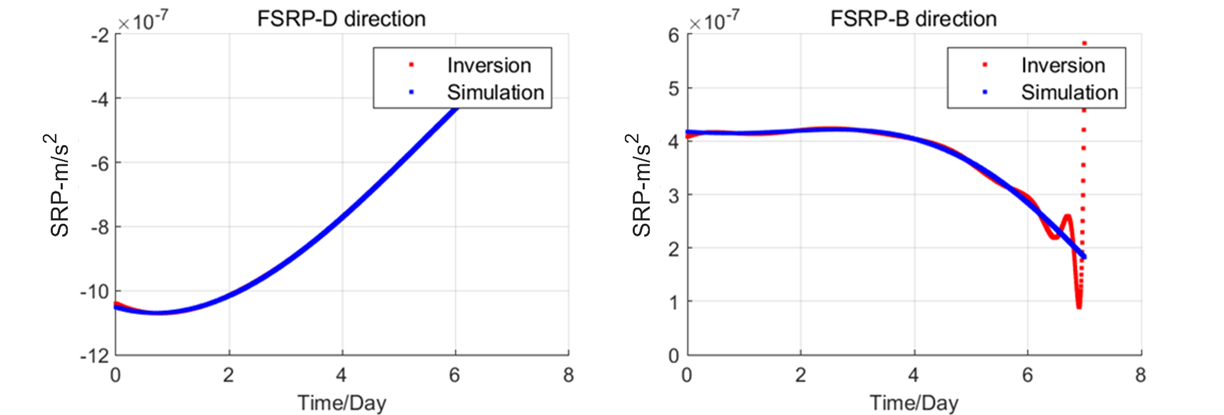}}
    \caption{OD results using the general ENCM with a 7‑day data arc.}
    \label{fig:l4_7day}
\end{figure}

\begin{figure}[htp]
    \centering
    \subfigure[OD position error.]{\includegraphics[width=4.5cm]{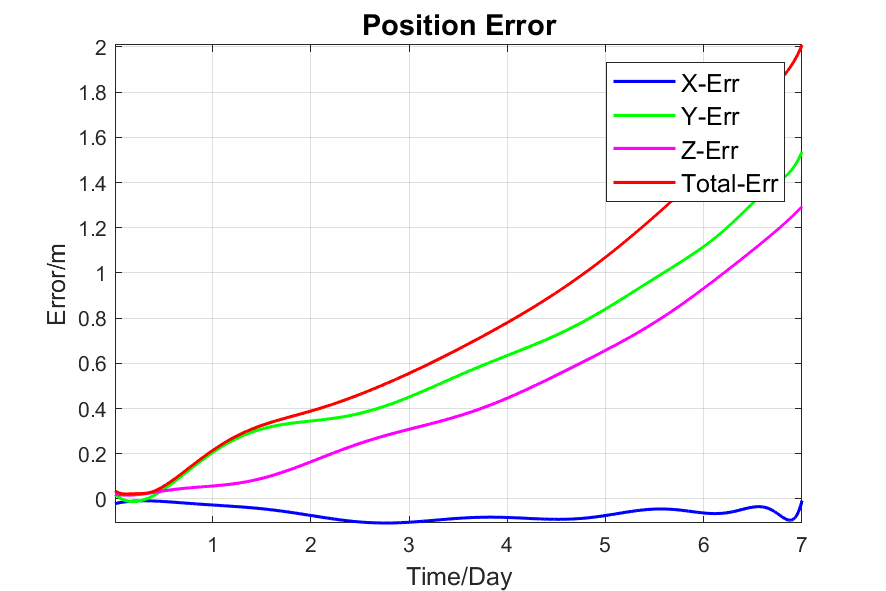}}
    \subfigure[SRP estimated with the Sub-ENCM.]{\includegraphics[width=9cm]{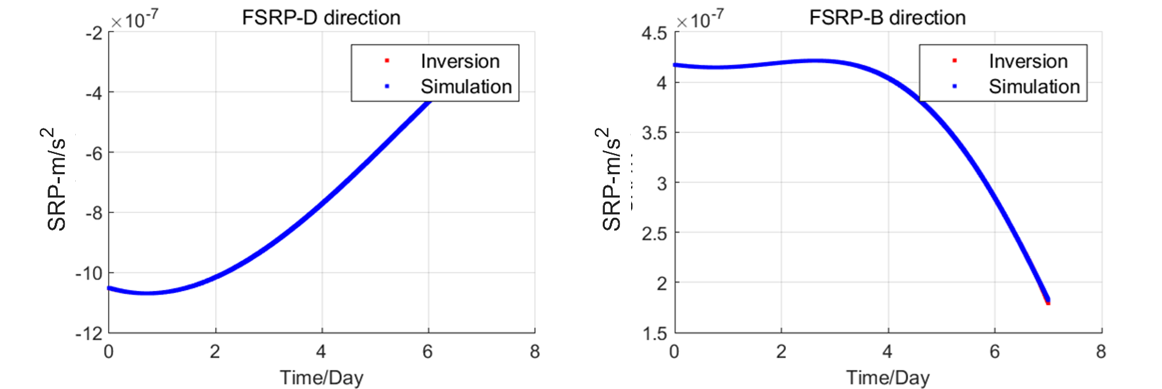}}
    \caption{OD results using the Sub-ENCM with a 7‑day data arc.}
    \label{fig:subencm_7day}
\end{figure}

\subsection{The Problem of NRHOs}
\par Compared to other types of orbits, NRHOs have some remarkable characteristics. NRHOs belong to the halo orbit family \cite{zimovan_2020}, yet they can also be approximated as highly eccentric lunar orbits. The north NRHO used in this study is shown in Fig.~14. It features a very small perilune distance and correspondingly large eccentricity. This leads to a very high sensitivity of orbit propagation errors to the force model errors. Even with the ENCM, a perfect match to the box-wing reference acceleration is not achieved. This residual modeling error leads to an intriguing phenomenon: OD errors remain acceptable as long as the perilune passage is excluded from the data arc, but they increase significantly when the perilune is included. This sensitivity is precisely why a 4-day arc (avoiding perilune) was selected for Tests~3 and 6 in Section~4.

\begin{figure}
    \centering
    \includegraphics[width=12cm]{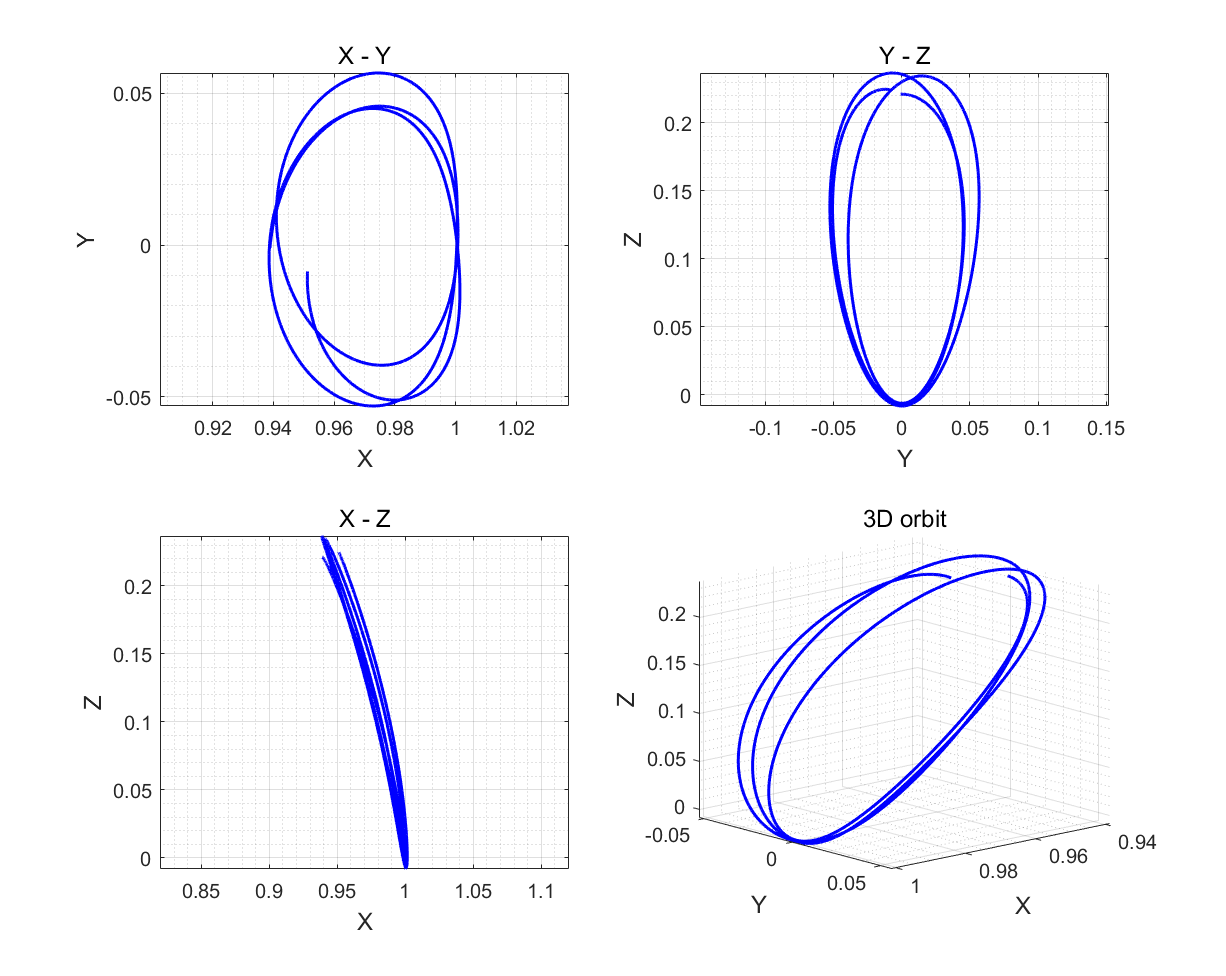}
    \caption{Example of a north NRHO in the Earth–Moon synodic coordinate system.}
\end{figure}

\subsection{Other Attitude Configurations}
\par In Section~3.1, the satellite attitude definition used in this work is outlined. In practice, however, satellites may operate under different attitude profiles. As an illustrative case, consider a constant yaw offset between the actual body-fixed +$\mathbf{X}'$ axis and the nominal +$\mathbf{X}$ axis defined in Section~3.1. This offset can be described by two constant angles.
\par In such a case, for the $\mathbf{D}$ and $\mathbf{B}$ directions, such a yaw offset merely introduces a constant phase shift in the arguments of Eq.~(23), leaving the frequency spectrum unchanged. In contrast, the $\mathbf{Y}$ component is affected differently. In the baseline models, $F_Y$ is modeled as a constant. The yaw offset, however, transforms part of the SRP acceleration from the $\mathbf{B}$ direction into the $\mathbf{Y}$ direction. Consequently, the expression for $F_Y$ becomes analogous to that for $F_B$, taking the form:
\begin{equation}
\begin{aligned}
{F}_{Y}&= Y_1+Y_2\cos{u_m}+Y_3\sin{u_m}+Y_4\cos{3u_m}+Y_5\sin{3u_m}\\
    & \quad +Y_6\cos{(u_p+u_m)}+Y_7\sin{(u_p+u_m)}+Y_8\cos{3(u_p+u_m)}+Y_9\sin{3(u_p+u_m)}\\
    & \quad +Y_{10}\cos{5(u_p+u_m)}+Y_{11}\sin{5(u_p+u_m)}+Y_{12}\cos{(u_p-u_m)}\\
    & \quad +Y_{13}\sin{(u_p-u_m)}+Y_{14}\cos{(2u_p+u_m)}+Y_{15}\sin{(2u_p+u_m)}.
\end{aligned}
\end{equation}

\section{Conclusion}
\par Objects in the cislunar space have recently attracted increasing attention. For most such objects, solar radiation pressure (SRP) is a significant perturbation, making an accurate SRP model crucial for high-accuracy OD. Currently, there is no suitable empirical SRP model for satellites in the cislunar space. This paper addresses the problem by developing a new empirical SRP model (termed ENCM) suitable for common cislunar orbits under specific attitude assumptions. Unlike the ECOM, the ENCM is related to two fundamental angles: the angle $u_m$ between the Earth-Moon line and Sun-Earth line, and the angle $u_p$ between the satellite’s radius vector from its orbital center and the projection of the Earth–Moon line onto the satellite’s orbital plane within the synodic frame.
\par This paper first derives specific sub‑models for several distinct orbital scenarios and synthesizes them into a general empirical model (general ENCM) for cislunar satellites. The validity of the ENCM is firstly verified by comparing it against reference SRP accelerations generated by the box‑wing model. Subsequently, the ENCM is integrated into the OD process. Comparisons of OD results obtained with the ENCM against those from the cannonball and ECOM models demonstrate its significantly superior performance. This confirms the unique advantage of the ENCM for OD of cislunar objects, recommending its adoption in future applications. The general ENCM contains many coefficients and accurate determination of these coefficients requires a longer arc and more observation data. However, many of these coefficients are intrinsically zero in specific scenarios; retaining them in the general model can introduce spurious estimation errors. Therefore, when the operational scenario is known, the corresponding sub‑models (sub‑ENCMs) are preferable. Numerical tests confirm superiority of  the sub‑ENCMs over the general ENCM, particularly for short data arcs. A brief discussion is also provided for cases where the satellite’s primary pointing direction deviates from the assumed configuration, noting that only the $\mathbf{Y}$‑component expression requires modification while the fundamental angular combinations remain unchanged. In summary, the ENCM is more accurate than the cannonball model and the ECOM for cislunar targets, and it is highly recommended.

\newpage
\appendix
\section{Initial orbital parameters}
\numberwithin{equation}{section}
\setcounter{equation}{0}
\numberwithin{figure}{section}
\setcounter{figure}{0}
\numberwithin{table}{section}
\setcounter{table}{0}
\par Table A.1 displays the initial positions and velocities used in this paper. All of these orbits are in the Earth-centered Celestial Reference System at 2028-01-01-00-00-000 (TDB. The difference between DRO-Small and DRO-Large lies in their sizes with respect to the Moon.) 

\begin{table}[htp]
    \renewcommand\arraystretch{1.2}
    \caption{\textbf{Initial positions and velocities of the orbits used in this study in the Earth-centered Celestial Reference System}}
    \centering
    \small
    \begin{tabular}{|c|c|c|} \hline
    \textbf{Orbit Types} & \textbf{Position (x,y,z) [m]} & \textbf{Velocity (x,y,z) [m$\cdot$s$^{-1}$]} \\ \hline
    DRO Small & 313304695.22\quad-217821887.04\quad-79977681.01   & 539.30\quad114.30\quad609.82\\ \hline
    DRO Large & 113564142.90\quad-26564827.15\quad-120219404.61 & 1036.90\quad560.71\quad351.89\\ \hline
    NRHO North & 305447851.33\quad-244389594.50\quad643794991.94 & 584.78\quad724.19\quad411.30 \\ \hline
    L4 Orbit & 306098727.61\quad160361769.95\quad265453462.73 & -438.89\quad722.91\quad459.05\\ \hline
    IGSO-1 & 42164137.00 \quad 0.000 \quad 0.000 & 0.000\quad1763.55\quad2518.62 \\ \hline
    IGSO-2 & -21082068.50\quad-20944266.19\quad-29911512.01 & 2662.73\quad-881.778\quad-1259.31 \\ \hline
    IGSO-3 & -21082068.50\quad20944266.19\quad29911512.01 & -2662.73\quad-881.77\quad-1259.31 \\ \hline
    \end{tabular}
    \label{}
\end{table}

\section{Expressions of Sub-ENCMs}
\numberwithin{equation}{section}
\setcounter{equation}{0}
\numberwithin{figure}{section}
\setcounter{figure}{0}
\numberwithin{table}{section}
\setcounter{table}{0}

\begin{itemize}
    \item [(1)] the sub-model for Scenario 2 (termed ENCM-2)
\begin{equation}
\begin{aligned}\left\{\begin{aligned}
    {F}_{D}&= D_1+D_2\cos{2u_m}+D_3\sin{2u_m}+D_4\cos{4u_m}+D_5\sin{4u_m}\\
    {F}_{Y}&= Y_1\\
    {F}_{B}&= B_1+B_2\cos{u_m}+B_3\sin{u_m}+B_4\cos{3u_m}+B_5\sin{3u_m}
\end{aligned}\right.\end{aligned}.
\end{equation}

    \item [(2)] the sub-model for Scenario 3 (termed ENCM-3)
\begin{equation}
\begin{aligned}\left\{\begin{aligned}
    {F}_{D}&= D_1+D_2\cos{2(u_p+u_m)}+D_3\sin{2(u_p+u_m)}+D_4\cos{2(u_p-u_m)}\\
    & \quad +D_5\sin{2(u_p-u_m)}+D_6\cos{2u_p}+D_7\sin{2u_p}+D_8\cos{2u_m}+D_9\sin{2u_m}\\
    {F}_{Y}&= Y_1\\
    {F}_{B}&= B_1+B_2\cos{(u_p+u_m)}+B_3\sin{(u_p+u_m)}+B_4\cos{3(u_p+u_m)}+B_5\sin{3(u_p+u_m)}\\
    & \quad +B_6\cos{5(u_p+u_m)}+B_7\sin{5(u_p+u_m)}+B_8\cos{(u_p-u_m)}+B_9\sin{(u_p-u_m)}
\end{aligned}\right.\end{aligned}.
\end{equation}
    
    \item [(3)] the sub-model for Scenario 4 (termed ENCM-4)
\begin{equation}
\begin{aligned}\left\{\begin{aligned}
    {F}_{D}&= D_1+D_2\cos{2(u_p+u_m)}+D_3\sin{2(u_p+u_m)}+D_4\cos{4(u_p+u_m)}+D_5\sin{4(u_p+u_m)}\\
    & +D_6\cos{(u_p+2u_m)}+D_7\sin{(u_p+2u_m)}+D_8\cos{(3u_p+2u_m)}+D_9\sin{(3u_p+2u_m)}\\
    {F}_{Y}&= Y_1\\
    {F}_{B}&= B_1+B_2\cos{(u_p+u_m)}+B_3\sin{(u_p+u_m)}+B_4\cos{3(u_p+u_m)}+B_5\sin{3(u_p+u_m)}\\
    & +B_6\cos{5(u_p+u_m)}+B_7\sin{5(u_p+u_m)}+B_8\cos{(2u_p+u_m)}\\
    & +B_9\sin{(2u_p+u_m)}+B_{10}\cos{u_m}+B_{11}\sin{u_m}
\end{aligned}\right.\end{aligned}.
\end{equation}
    
    \item [(4)] the sub-model for Scenario 5 (termed ENCM-5)
\begin{equation}
\begin{aligned}\left\{\begin{aligned}
    {F}_{D}&= D_1+D_2\cos{2u_m}+D_3\sin{2u_m}+D_4\cos{4u_m}+D_5\sin{4u_m}\\
    & +D_6\cos{(u_p-2u_m)}+D_7\sin{(u_p-2u_m)}+D_8\cos{(u_p+2u_m)}+D_9\sin{(u_p+2u_m)}\\
    {F}_{Y}&= Y_1\\
    {F}_{B}&= B_1+B_2\cos{u_m}+B_3\sin{u_m}+B_4\cos{3u_m}\\
    & +B_6\cos{(u_p-u_m)}+B_7\sin{(u_p-u_m)}+B_{8}\cos{(u_p+u_m)}+B_{9}\sin{(u_p+u_m)}
\end{aligned}\right.\end{aligned}.
\end{equation}
    
\end{itemize}

\nocite{*}

\bibliographystyle{plain}   
\bibliography{refs}  

\end{document}